\documentclass[aps,prb,reprint,twocolumn,superscriptaddress,PhySH,showkeys,floatfix,nobibnotes]{revtex4-2}

\usepackage{graphicx}
\usepackage{epstopdf}
\usepackage{amssymb}
\usepackage{amsmath}
\usepackage{bm}
\usepackage{braket}
\usepackage{color}
\usepackage{float}

\usepackage[colorlinks=true,linkcolor=blue,citecolor=blue,urlcolor=blue]{hyperref}

\begin{document}
	
	\title{Dissipation-induced Nonlinear Topological Gear Switching}
	
	\author{Xuzhen Cao}
	\thanks{These authors contribute equally to this work.}
	\affiliation{State Key Laboratory of Quantum Optics Technologies and Devices, Institute of Laser Spectroscopy, Shanxi University, Taiyuan 030006, China} 
	\affiliation{Collaborative Innovation Center of Extreme Optics, Shanxi University, Taiyuan 030006, China}
	\affiliation{School of Computer and Information Technology, Shanxi University, Taiyuan 030006, China}
	
	\author{Xiaolin Li}
	\thanks{These authors contribute equally to this work.}
	\affiliation{School of Physics, Northwest University, Xi'an 710127, China}
	\affiliation{Graduate School of China Academy of Engineering Physics, Beijing 100193, China}
	
	\author{Liang Bai}
	\affiliation{Key Laboratory of Computational Intelligence and Chinese Information Processing of Ministry of Education, Institute of Intelligent Information Processing, Shanxi University, Taiyuan 030006, China}
	
	\author{Zhaoxin Liang}
	\email{zhxliang@zjnu.edu.cn}
	\affiliation{Department of Physics, Zhejiang Normal University, Jinhua 321004, China}
	
	\author{Li-Chen Zhao}
	\email{zhaolichen3@nwu.edu.cn}
	\affiliation{School of Physics, Northwest University, Xi'an 710127, China}
	\affiliation{Shaanxi Key Laboratory for Theoretical Physics Frontiers, Xi'an 710127, China}
	\affiliation{NSFC-SPTP Peng Huanwu Center for Fundamental Theory, Xi'an 710127 and Hefei 230026, China}
	
	\author{Ying Hu}
	\email{huying@sxu.edu.cn}
	\affiliation{State Key Laboratory of Quantum Optics Technologies and Devices, Institute of Laser Spectroscopy, Shanxi University, Taiyuan 030006, China} 
	\affiliation{Collaborative Innovation Center of Extreme Optics, Shanxi University, Taiyuan 030006, China}
	
	\begin{abstract} 
		Nonlinear interaction enables topological phenomena impossible in linear systems. A paradigm is nonlinear Thouless pump, where the transport of solitons can be topologically quantized even when band occupation is nonuniform. Such nonlinear quantization traditionally requires a time-periodic Hamiltonian with static nonlinearity and, much as in the linear case, is inherently independent of pumping speed. Instead, we demonstrate a dissipation-induced topological gear switching, where quantized soliton transport can be switched on and off via the adiabatic pumping speed itself. This phenomenon has no counterpart in prior conservative nonlinear pumps, nor in linear non-Hermitian pumps. Crucially, quantization here no longer requires a time-periodic nonlinear Hamiltonian; it stems from a genuinely non-equilibrium mechanism captured by an effective conservative model whose \textit{nonlinearity varies aperiodically in time}. Remarkably, a quantized nonlinear transport can be induced even when this nonlinear aperiodic driving is such that the system is pumped from the linear to nonlinear regimes. Our results open a route toward nonequilibrium nonlinear topological matter, where topological effect is dynamically reconfigurable via time-varying nonlinearities, with experimental implications for photonic, atomic, or superconducting platforms and beyond.
	\end{abstract}
	
	\maketitle
	
	\section{Introduction}\label{Sec:Intro}
	
	Interactions or nonlinearities can essentially modify conventional topological paradigms~\cite{Smirnova2020,Citro2023,Szameit2024,Leykam2026,Chen2025}, leading to effects with no linear counterpart. A cornerstone is nonlinear Thouless pumping~\cite{Jurgensen2021}, where the transport of a soliton - a stable nonlinear excitation - is topologically quantized even when the underlying band is non-uniformly occupied, violating a basic condition of the linear quantization paradigm. This nonlinear quantization relies instead on a time-periodic Hamiltonian with static nonlinearity, so that the instantaneous stable soliton returns to itself (up to translation) after each cycle. Consequently, a nonlinearly induced topological phase transition arises which, importantly, does not depend on the adiabatic pumping speed, much as in the linear case~\cite{Thouless1982,Thouless1983,Di2010}. To date, nonlinear topological physics has blossomed into an exciting frontier spanning photonics, cold atoms to condensed matter~\cite{Bongiovanni2021,Walter2023,Viebahn2024, Jurgensen2021,Zhu2025}, yet the focus remains primarily on conservative systems and static nonlinearity~\cite{Jurgensen2022,Mostaan2022,Qidong2022,Fu2022,Jurgensen2023,Aligia2023,Tuloup2023,Cao2024,Lyu2024,Cao2025,Tao2025,Bai2025a,Bai2025b}. 
	
	\begin{figure*}
		\centering
		\includegraphics[width=0.9\textwidth]{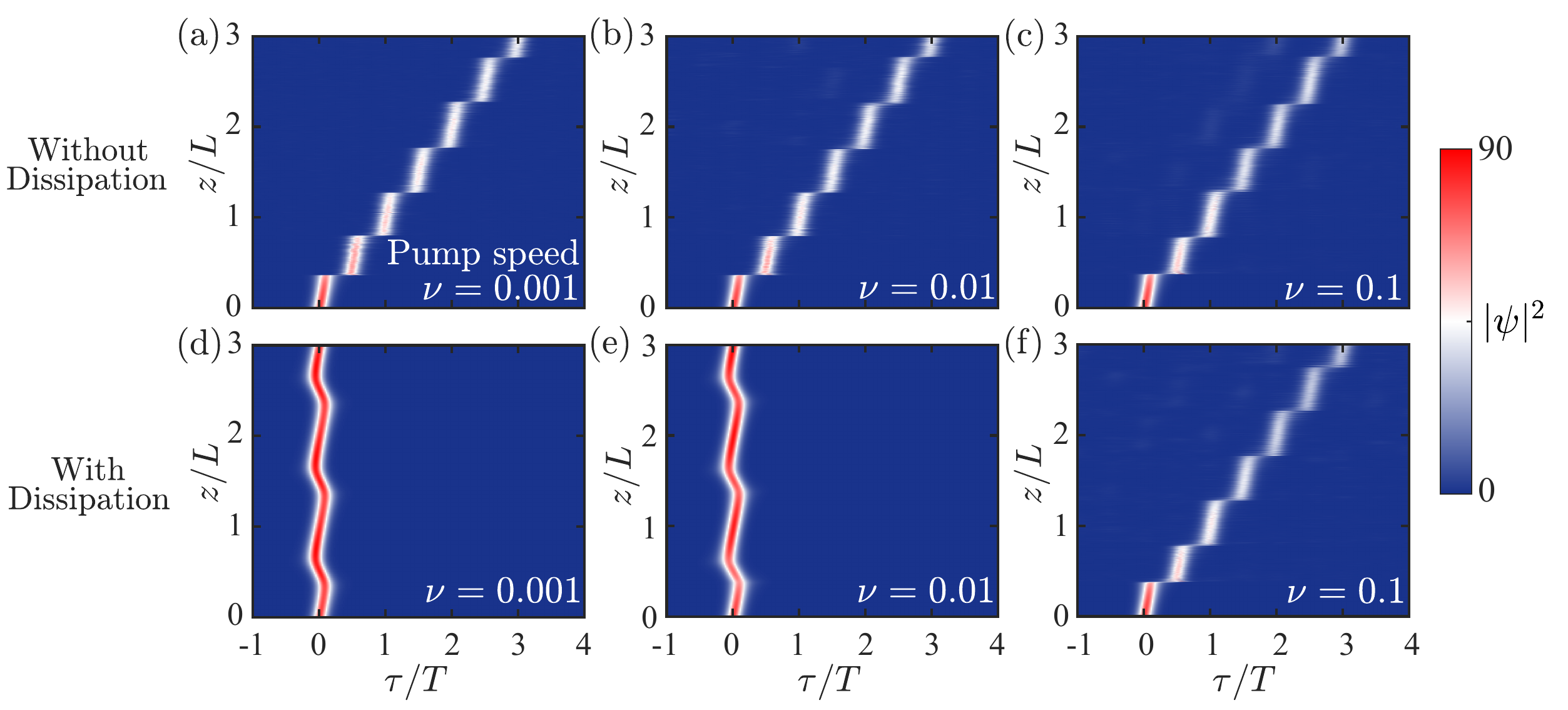}
		\caption{Dissipatively induced topological gear switching. Propagation of same input in (a)-(c) the conservative ($\delta=\beta=\epsilon=0$) and (d)-(f) weakly dissipative regimes ($\delta = 9 \times 10^{-4}$, $\beta = 10^{-5}$, $\epsilon = -10^{-5}$), for slow pump speeds $\nu=0.001,0.01,0.1$. Numerical results are obtained by solving Eq.~(\ref{CGLE}) with $g = 1$ and $V_s = V_l = 25$. The initial soliton is taken as the instantaneous nonlinear eigenstate of $H_\textrm{NL}(0)$ with power $P(0) = 15.2$. }\label{Fig1}
	\end{figure*}
	
	Recent experimental advances in synthetic systems enable access to regimes where dissipation (non-Hermiticity) and nonlinearity are combined~\cite{Konotop2016,Jeon2020,Xia2021,Liu2021,Pernet2022,Ravets2025}. This motivates a crucial question as to whether their interplay induces qualitatively novel topological scenarios. Preliminary studies have revealed intriguing possibilities, demonstrating for instance topological lasing~\cite{St-Jean2017,Miguel2018,Gal2018}, topological gap solitons in driven-dissipative plaritons~\cite{Pernet2022}, and novel bifurcation in nonlinear non-Hermitian systems~\cite{Kawabata2025}. Here, we unveil a phenomenon with no prior analogue: a topological gear switching in a dissipative nonlinear Thouless pump. We show that a nonlinear topological transport can be switched on and off through the adiabatic pumping speed, in stark contrast to established paradigms where it is irrelevant to the outcome.
	
	This topological gear switching cannot be explained within the conventional equilibrium framework~\cite{Jurgensen2021,Jurgensen2022,Mostaan2022,Qidong2022,Fu2022,Jurgensen2023,Aligia2023,Tuloup2023,Cao2024,Lyu2024,Cao2025,Tao2025,Bai2025a,Bai2025b,Bongiovanni2021,Walter2023,Viebahn2024}, where quantization follows from adiabatically tracking nonlinear eigenstates of a time-periodic Hamiltonian with static nonlinearity. Instead, it originates from a  genuinely non-equilibrium mechanism that does not require a time-periodic nonlinear Hamiltonian. In particular, we show that the combined interplay of dissipation, nonlinearity and pump is captured by an effective conservative model \textit{whose nonlinearity depends aperiodically in time}. Crucially, it is the excitation of this aperiodic Hamiltonian, instead of the original periodic one, that governs the quantized nonlinear pumping. Remarkably, a quantized shift is induced even when the nonlinear driving is strongly aperiodic, making the system transitions from a linear to nonlinear regimes. Our results open a new route toward non-equilibrium nonlinear topological matter, with implications for dynamically configurable nonlinear topological effects. Our findings are feasible in generic experimental platforms ranging from photonic time crystals to driven-dissipative simulators.
	
	\section{Speed-dependent topological pumping by dissipation}\label{Sec:Model}
	We illustrate the essential physics based on the light propagation in a dissipative photonic waveguide generically governed by the dimensionless equation~\cite{Aranson2002}
	\begin{equation}\label{CGLE}
		i\frac{\partial \psi}{\partial z} =[H_0(z) - g|\psi|^2]\psi+ i\mathcal{D}(\psi)\psi
	\end{equation}
	where $H_0(z)\equiv -\frac{1}{2}\partial_\tau^2+V_{\text{p}}(z,\tau)$ is the linear Hamiltonian, and $\mathcal{D}(\psi) = \beta\partial_\tau^2 + \delta + \epsilon|\psi|^2$ is the dissipator. Here, $\psi$ is the wavefunction (envelope of the electric field), $z$ is the propagation distance, and $\tau$ is the retarded time. The potential 
	\begin{equation}
		V_{\text{p}}(z,\tau) = -V_s\cos^2(2\pi \tau) -V_l\cos^2(\pi \tau - \nu z) \label{OSL}
	\end{equation}
	represents a topological Thouless pump with amplitudes $V_{s,l}$, where the phase $\phi=\nu z$ is slowly varied at a speed $\nu$ (the pump period is $L=\pi/\nu$). This pump can be realized in photonic systems with state-of-the-art phase modulation techniques~\cite{Song2020,Andrea2022,Chen2021,Englebert2023,Mao2025,Copie2025}. The nonlinear term $-g|\psi|^2$ governs self-focusing ($g>0$) or defocusing ($g<0$) dynamics. The dissipator $\mathcal{D}$ combines linear loss ($\beta$), gain ($\delta$), and nonlinear loss or gain ($\epsilon$) effects. Without dissipation ($\mathcal{D}=0$), Eq.~(\ref{CGLE}) reduces to the nonlinear Schr\"{o}dinger equation that describes conventional and conservative nonlinear Thouless pumping~\cite{Jurgensen2022,Mostaan2022,Qidong2022,Fu2022,Jurgensen2023,Aligia2023,Tuloup2023,Cao2024,Lyu2024,Cao2025,Tao2025,Bai2025a,Bai2025b}, with the norm $P(z) = \int_{-\infty}^{+\infty} |\psi(z,\tau)|^2 d\tau= P(0)$ is conserved along the synthetic-time dimension $z$. Without the pump, Eq.~(\ref{CGLE}) underlies the description of dissipative solitons and dissipative nonlinear dynamics~\cite{Aranson2002}. For convenience, we label the nonlinear Hamiltonian $H_\textrm{NL}(z)\equiv H_0(z)-g|\psi|^2$.
	
	\begin{figure}
		\centering
		\includegraphics[width=1\columnwidth]{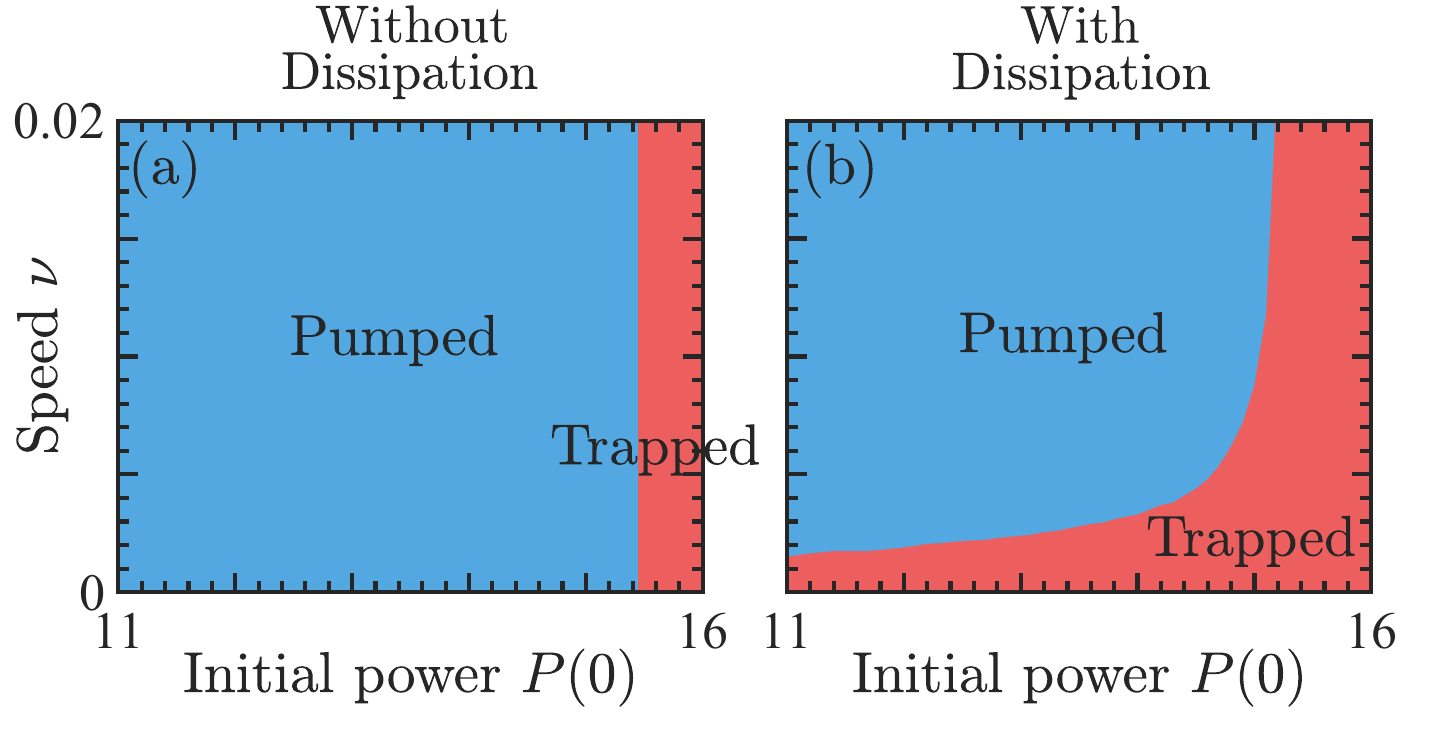}
		\caption{Phase diagrams in (a) conservative vs. (b) weakly dissipative regimes. Solitonic displacement after first cycle is scanned as a function of $P(0)$ and pump speed $\nu$; the displacement equals one unit (zero) in the pumped (trapped) phase. Results are obtained from Eq.~(\ref{CGLE}) with $g=1$, $V_{s,l}=25$, for $\beta$, $\delta$ and $\epsilon$ in Fig.~\ref{Fig1}. The initial wavefunction is a nonlinear eigenstate of $H_\textrm{NL}(0)$~\cite{Supple}. }\label{Fig2}
	\end{figure}
	
	We demonstrate the differences between
	conservative and dissipative nonlinear Thouless pumping under different adiabatic pump speeds $\nu$, by propagating the same input soliton in two different regimes: (i) a conservative regime ($\beta=\delta=\epsilon=0$) characterized by nonlinear strength $gP(0)$, as a reference for our study; (ii) a weakly dissipative regime with $\beta = 10^{-5}$, $\delta = 9 \times 10^{-4}$ and $\epsilon = -10^{-5}$. We numerically simulate Eq.~(\ref{CGLE}) with a 4th-order Runge-Kutta scheme~\cite{Supple}. For the initial state, we take a nonlinear eigenstate of $H_\textrm{NL}(0)$ - which bifurcates from the lowest band [Fig.~\ref{Fig3}(a)] of the linear Hamiltonian $H_0(0)$ - obtained via self-consistent iterative method~\cite{Supple}. In all simulations, we set $g=1$, $V_s=V_l=25$ without loss of generality. Figure~\ref{Fig1} demonstrates the evolution of an initial soliton with $P(0)=15.2$ over three pump cycles, for various speeds $\nu=0.001,0.01,0.1$ that ensures adiabatic evolution in the linear limit. (Details on the soliton's characteristics, including center-of-mass displacement quantified by $\langle\tau\rangle={\int_{-\infty}^{+\infty} \tau|\psi|^2 d\tau}/{P(0)}$ and a higher order norm~\cite{Jurgensen2023,Supple}, as well as participation ratio, are provided in Supplementary Material.) 
	
	In the conservative regime [Figs.~\ref{Fig1}(a)-(c)], the soliton is pumped by one unit independent of $\nu$, as in previous works~\cite{Jurgensen2021,Qidong2022}. In stark contrast, the weakly dissipative scenario [Figs.~\ref{Fig1}(d)-(f)] reveals a dramatically different, speed-dependent fate for the soliton. For $\nu=0.001,0.01$ [Figs.~\ref{Fig1}(d)-(e)], the input soliton becomes dynamically trapped near its initial position, yielding zero displacement after a full cycle. This trapping is counterintuitive because the gain and loss are exceptionally weak ($\beta=10^{-5}$, $\delta=9\times10^{-4}$, $\epsilon P(0)=-1.5\times10^{-4}$), so that one might expect the dynamics to closely approach their conservative counterpart [Figs.~\ref{Fig1}(a)-(b)]. Instead, quantized shift [Fig.~\ref{Fig1}(f)] occurs only for the adiabatic speed $\nu$ above a threshold. 
	
	This speed-dependent nonlinear topological pumping, induced by weak dissipation, is supported by the phase diagrams in Fig.~\ref{Fig2}. Here, the soliton's center-of-mass displacement at the end of first cycle is scanned as a function of $P(0)$ and speed $\nu$~\cite{Supple}, for parameters in Fig.~\ref{Fig1}. In the conservative case [Fig.~\ref{Fig2}(a)],  tuning the nonlinearity $gP(0)$ drives a topological phase transition of solitons between a trapped phase and quantized displacement by one unit as dictated by the Chern number of the underlying linear band [Fig.~\ref{Fig3}(a)], irrelevant of $\nu$. By contrast, the phase transition in the dissipative regime [Fig.~\ref{Fig2}(b)] strongly varies with $\nu$ [Fig.~\ref{Fig2}(b)], with a slower pumping enhancing trapping. Phase diagrams for different dissipative parameters are presented in Supplementary Material. These results demonstrate a speed-dependent nonlinear topological phase transition - termed topological gear switching - that has no direct analogue in conventional conservative nonlinear Thouless pumping. 
	
	\section{Physical picture}\label{Sec:principle}
	To understand solitonic motion in the weakly dissipative regime, we first recall the mechanism underlying conventional conservative nonlinear Thouless pumping~\cite{Jurgensen2022}. There, the quantized solitonic displacement arises from adiabatic following of the instantaneous stable nonlinear eigensolution of $H_\textrm{NL}(\phi)\varphi_{c}(\phi)=E_c(\phi)\varphi_{c}(\phi)$ at each pump phase $\phi=\nu z$ [Eq.~(\ref{OSL})]~\cite{Supple}. Because $H_\textrm{NL}(\phi)$ is periodic in $\phi$, $\varphi_{c}$ returns to itself at the end of each cycle, up to translation invariance, and hence quantization, independent of $\nu$. A natural extension of this mechanism to the weakly dissipative regime ($\mathcal{D}\neq 0$) would suggest that the displacement follows the instantaneous stable dissipative soliton solution $\varphi_{d}(\phi)$ of the non-Hermitian operator $\mathcal{H}_\textrm{NH}(\phi)\equiv H_\textrm{NL}(\phi)+i\mathcal{D}$ parameterized by $\phi$. Stabilizing a dissipative soliton requires double balances: between nonlinearity and dispersion, and between gain and loss. This corresponds to a real-eigenvalue solution $\mathcal{H}_\textrm{NH}(\phi)\varphi_{d}=E_d\varphi_{d}$ with $E_d\in \mathbb{R}$~\cite{Supple}. Since $\mathcal{H}_\textrm{NH}$ is also $\phi$ periodic, this extension would likewise predict $\nu$-independent quantization. The observed speed-dependent topological gear switching in Figs.~\ref{Fig1} and \ref{Fig2} directly refutes this conjecture. This signals the breakdown of the static eigenstate-following paradigm and suggests a \textit{dynamical} mechanism absent in conventional Thouless pumping, where the pump speed itself acts as a topological control parameter.
	
	Indeed, topological gear switching can be explained within a Born-Oppenheimer-type picture that separates pumping dynamics into fast and slow timescales. For weak dissipation considered here, the nonlinear-dispersion energy balance is rapidly established on a short timescale $z_c\sim [\textrm{max}(V_s, V_l)]^{-1}$ to set the soliton's profile, while the gain-loss balance - which determines the soliton's intensity - occurs on a much longer timescale $z_d\sim [\textrm{max}(|\epsilon| P(0), |\beta|,|\delta|)]^{-1}$; for instance,  $z_c\sim 10^{-2}$ and $z_d\sim 10^{4}$ for Fig.~\ref{Fig1}. Thus, for a pump in the regime $z_c\ll \nu^{-1} \ll z_d$, the soliton’s profile adjusts almost instantaneously to the slowly varying intensity induced by gain and loss. 
	
	\begin{figure}
		\centering
		\includegraphics[width=1\columnwidth]{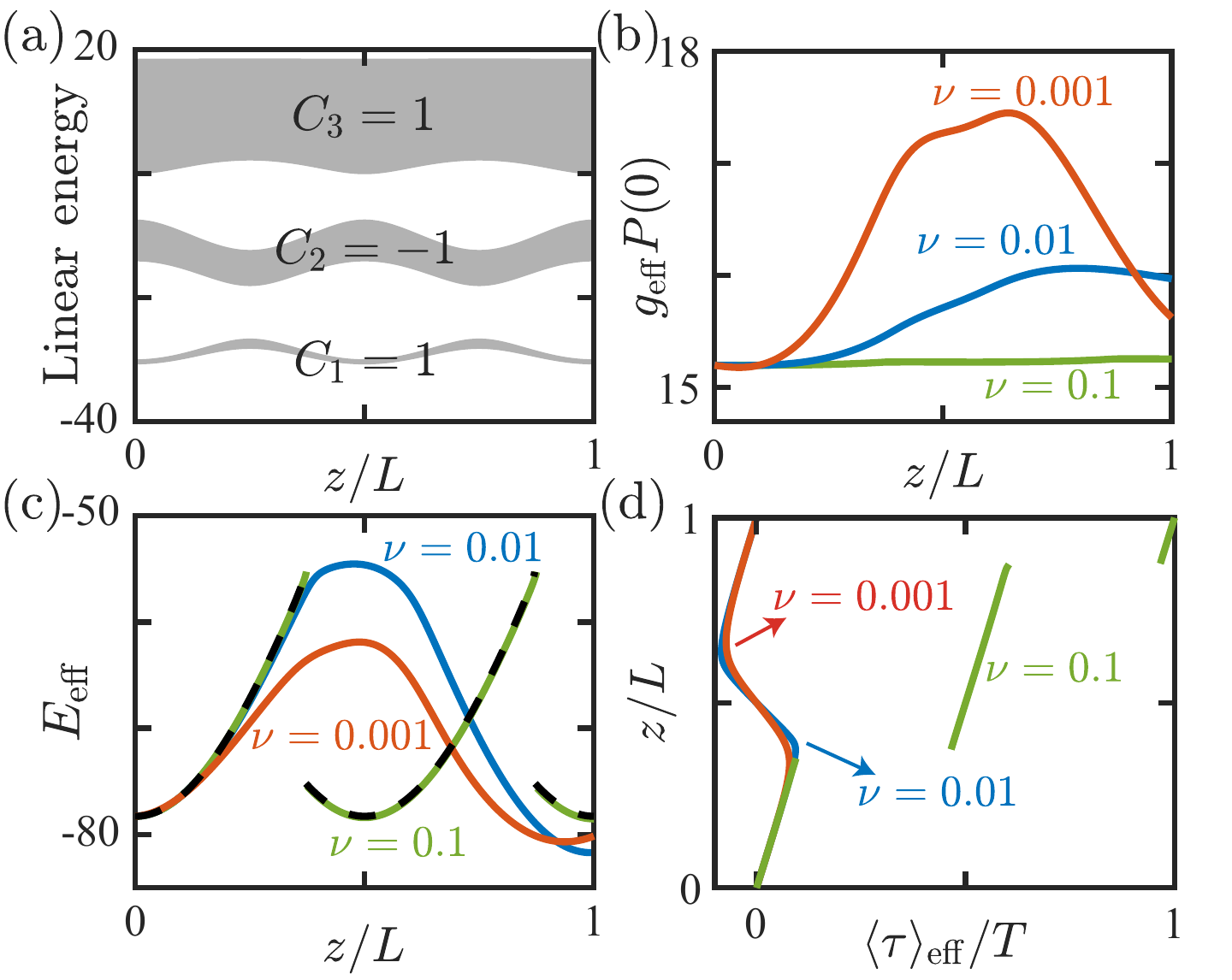}
		\caption{Mechanism of topological gear switching. (a) The three lowest linear energy bands and their Chern numbers in the linear, conservative regime. (b) Time-varying effective nonlinearity $g_{\text{eff}}(\tilde{z}) P(0)$ in Eq.~(\ref{effective}). (c) Energy $E_{\text{eff}}(z)$ and (d) center-of-mass position $\langle \tau \rangle_{\mathrm{eff}}$ of the effective instantaneous nonlinear eigenstate $\varphi_{\mathrm{eff}}(z)$. In (c), black curve denotes the dissipation-free counterpart as the reference. Other parameters are same as Figs.~\ref{Fig1}(d))-(f). 
		}\label{Fig3}
	\end{figure}
	
	The above Born-Oppenheimer-type picture motivates an ansatz for solutions of Eq.~(\ref{CGLE}) in the form 
	\begin{equation}\label{ansatz}
		\psi(z,\tau)\approx f(z)\varphi_{\mathrm{eff}}(z,\tau), 
	\end{equation}
	with $f(0)=1$. Here, the wavefunction $\varphi_{\mathrm{eff}}(z,\tau)$ evolves on the fast scale $z_c$, preserving its initial norm $\int_{-\infty}^{+\infty} d\tau \, |\varphi_{\mathrm{eff}}(z,\tau)|^2 = P(0)$ at each time-$z$, while $f(z)$ evolves on the slow scale $z_d$, capturing the dissipation-induced norm modulation. Substituting the ansatz~(\ref{ansatz}) into Eq.~(\ref{CGLE}) and rescaling $\tilde{z}=\nu z/\pi$, we derive an effective nonlinear Schr\"{o}dinger equation on the fast scale:
	\begin{equation}
		\label{effective}
		\left[H_0(\tilde{z}) - g_\textrm{eff}(\tilde{z})|\varphi_{\mathrm{eff}}|^2\right]\varphi_{\mathrm{eff}}= E_\textrm{eff}(\tilde{z})\varphi_{\mathrm{eff}};
	\end{equation}
	with $g_\textrm{eff}(\tilde{z}) = g|f(\tilde{z})|^2$. On the slow scale, the function $f(\tilde{z})$ evolves according to 
	\begin{equation}
		\label{f}
		\frac{\partial f}{\partial \tilde{z}} = \pi\frac{\gamma(\tilde{z})}{\nu} f; \quad 
		\gamma(\tilde{z}) \equiv \frac{\left\langle \varphi_{\mathrm{eff}}(\tilde{z})\left|\mathcal{D}\right|\varphi_{\mathrm{eff}}(\tilde{z})\right\rangle}{P(0)}.
	\end{equation}
	Detailed validation of the ansatz~(\ref{ansatz}) is presented in Supplementary Material.
	
	Equation (\ref{effective}) constitutes a central result of this work that captures the essential physics of topological gear switching. It shows that weakly non-Hermitian nonlinear pumping can be understood via an ``effective Hermitian model with an \textit{aperiodically time-varying} nonlinearity $g_\textrm{eff}(\tilde{z})$'', with a variation rate controlled by $\gamma/\nu$.  As illustrated in Fig.~\ref{Fig3}(b) for dissipative parameters in Fig.~\ref{Fig1}, when $\nu=0.001, 0.01$, the corresponding effective nonlinearity varies substantially over the first cycle, with an overall increase after the first period. For higher speed $\nu=0.1$, $g_\textrm{eff}(\tilde{z})\approx g$ with strongly suppressed modulation. This direct link between the pump speed and the time-dependent nonlinearity reflects an intrinsically non-equilibrium nature of the instantaneous soliton $\varphi_{\mathrm{eff}}$: although a local balance between nonlinearity and dispersion is rapidly established, the gain and loss do not equilibrate which, in turn, feeds back via the effective nonlinearity to modify the instantaneous solitary profile. 
	
	\begin{figure}
		\centering
		\includegraphics[width=1\columnwidth]{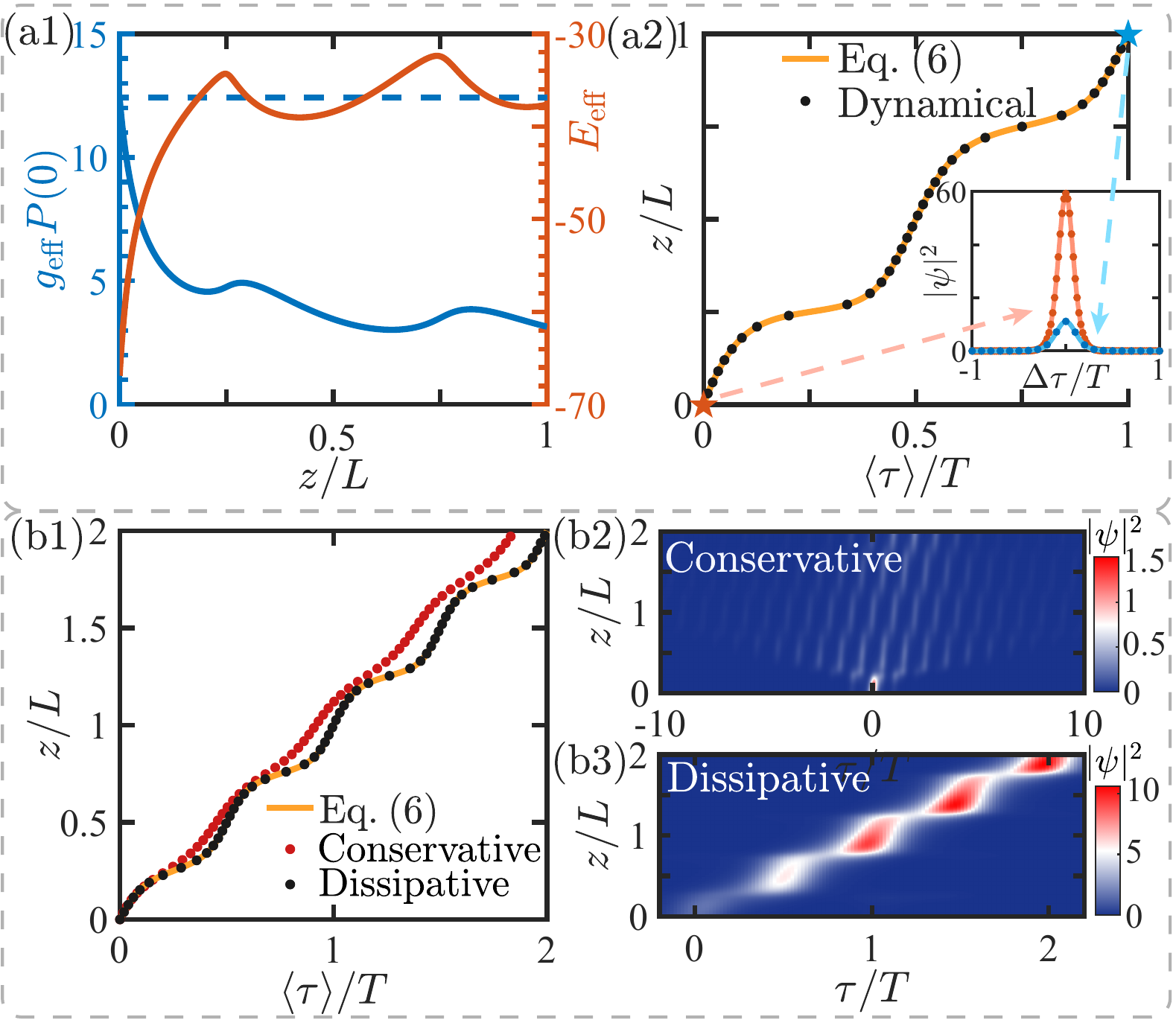}
		\caption{Quantized nonlinear pumping despite strongly aperiodic nonlinearity. (a1)-(a2) Quantization despite significantly deformed solitons after a cycle. Starting from a soliton ($P(0) = 12.42$), the system evolves under parameters $\delta = 10^{-3}$, $\beta = 5 \times 10^{-5}$, $\epsilon = -5 \times 10^{-5}$ and $\nu = 0.001$. (a1) $g_{\mathrm{eff}}(z) P(0)$ (blue) and $E_{\mathrm{eff}}(z)$ (red). Dashed line indicates dissipation-free counterpart. (a2) Although the soliton profile is significantly deformed after a full cycle (inset), its center of mass (black points, from Eq.~(\ref{CGLE})) adiabatically follows the effective instantaneous soliton $\varphi_{\mathrm{eff}}(\tilde{z})$ (yellow) and exhibits quantized displacement. (b1)-(b3) Quantization despite transformation from nonsolitary to solitary solutions. For a weak bare nonlinearity $P(0) = 0.5$, the system initializes in a linear pump regime.  In the conservative case, (b2) the initial wave packet disperses, whereas (b3) a soliton is dynamically formed in weakly dissipative regime. (b1) Center-of-mass displacement is not quantized in conservative regime, but quantized in weakly dissipative regimes despite fundamental changes in the excitations.  Parameters: $\delta = 1.2 \times 10^{-3}$, $\beta = 6.5 \times 10^{-5}$, $\epsilon = -6.5 \times 10^{-5}$. In (b1), dotted curves denote results from Eq.~(\ref{CGLE}); yellow curve denotes center-of-mass position of $\varphi_{\mathrm{eff}}(\tilde{z})$. All panels: $V_s = V_l = 25$.}\label{Fig4}
	\end{figure}
	
	The topological gear switching [e.g. Figs.~\ref{Fig1}(d)-(f)] can be understood through the adiabatic following of the instantaneous effective nonlinear eigenstate $\varphi_\textrm{eff}(z)$ in Eq.~(\ref{effective}). Its effective nonlinear eigenvalue $E_\textrm{eff}(z)$ and positions of the centers of mass
	\begin{equation}\label{tf}
		\langle \tau \rangle_{\mathrm{eff}}=\frac{\int_{-\infty}^{+\infty} \tau|\varphi_{\mathrm{eff}}|^2 d\tau}{P(0)}
	\end{equation}
	at each $z$ are plotted in Figs.~\ref{Fig3}(c) and (d). For $\nu=0.001, 0.01$, the effective spectrum arising from dissipation-modified nonlinearity is continuous (thus ensures adiabaticity in nonlinear regime) and aperiodic [Fig.~\ref{Fig3}(c)], qualitatively different from the conservative case (black dashed curve). The path of center-of-mass of the effective soliton returns to its initial position after a full pump cycle [Fig.~\ref{Fig3}(d)], consistent with the dynamical trapping in Fig.~\ref{Fig1}(d). By contrast, for $\nu=0.1$ (green curves in Figs.~\ref{Fig3}(c) and (d)), where $g_\textrm{eff}(\tilde{z})\approx g$, the spectrum closely approaches the dissipation-free band, and the instantaneous center-of-mass $\langle \tau \rangle_{\mathrm{eff}}$ is displaced by one unit after one cycle, in agreement with Fig.~\ref{Fig1}(f). Note that in the asymptotic long-time limit~\cite{Supple}, the system generally approaches a dynamical equilibrium, where $g_\textrm{eff}(\tilde{z})$ develops periodic oscillation and $\gamma$ oscillates around $0$, indicating the soliton converges to a limit cycle near the attractor fixed by double balances.
	
	Thus, the speed-dependent nonlinear pumping in the weakly dissipative regime is governed by an effective Hermitian model with a time-varying nonlinearity controlled by $\gamma/\nu$, following its instantaneous solutions, rather than the original non-Hermitian operator $\mathcal{H}_\textrm{NH}(z)$. 
	
	\section{Quantization from aperiodic time-varying nonlinearity}\label{Sec:QWPNH}
	The transport captured by Eq.~(\ref{effective}) goes conceptually beyond the standard paradigm of nonlinear Thouless pumping~\cite{Jurgensen2022,Mostaan2022}, which relies on a time-periodic nonlinear Hamiltonian (with constant nonlinearity) and quantization originates from the instantaneous soliton being identical after each cycle up to translation invariance. In stark contrast, here the effective nonlinear Hamiltonian is aperiodic, yet quantized transport still arises.
	
	To further highlight this aspect, we examine a case where the effective nonlinearity $g_\textrm{eff}(z)$ and eigenvalue $E_\textrm{eff}(z)$ [Fig.~\ref{Fig4}(a1)] vary so strongly that the effective instantaneous soliton $\varphi_{\mathrm{eff}}$ at the beginning and end of a cycle substantially changes its profiles and cannot reproduce itself [Fig.~\ref{Fig4}(a2) inset]. Nevertheless, the solitonic displacement still adiabatically follows the center-of-mass position of $\varphi_{\mathrm{eff}}(\tilde{z})$ [Eq.~(\ref{tf})] and exhibits a quantized shift after one cycle [Fig.~\ref{Fig4}(a2)] (longer-time dynamics can be found in Supplementary Material). A more dramatic example is illustrated in Figs.~\ref{Fig4}(b1)-(b3), where the initial eigenstate is not even a soliton. Here, the bare nonlinearity $gP(0)=0.5$ places the system in the linear regime in the absence of dissipation: the initial wave packet simply disperses without soliton formation [Fig.~\ref{Fig4}(b2)], and its displacement does not quantize [Fig.~\ref{Fig4}(b1)] due to nonuniform band occupation in the linear regime~\cite{Jurgensen2022}. By contrast, in the presence of weak dissipation, a soliton dynamically forms during the pumping [Fig.~\ref{Fig4}(b3)]. Importantly, despite this fundamental change in the state's character, its center-of-mass displacement becomes quantized after one cycle [Fig.~\ref{Fig4}(b1)].
	
	\section{Conclusion}\label{Sec:Concluding}
	In summary, we have uncovered a speed-dependent topological gear switching in a dissipative nonlinear Thouless pump, where the adiabatic pumping speed - previously an irrelevant parameter - becomes an essential control knob that turns quantized nonlinear transport on or off. The phenomenon is captured by an effective conservative model whose nonlinearity varies aperiodic in time, leading to a genuinely non-equilibrium mechanism for quantized transport. Our findings can be generalized to a wide range of non-Hermitian nonlinear lattice systems. 
	
	Our work extends the paradigm of nonlinear topological pumping beyond the constraint of a time-periodic nonlinear Hamiltonian. Unlike established mechanisms, which relies on a time-periodic nonlinear Hamiltonian with static nonlinearity and an instantaneous soliton returning to itself after each cycle, we show that quantized transport exists even when the nonlinearity is time-dependent and strongly aperiodic. In our system, aperiodicity is induced by dissipation, but we emphasize the same mechanism should extend to purely conservative systems with appropriately engineered nonlinear driving, opening a new route to control topological transport~\cite{Supple}. In a broader context, our work extends the studies of nonlinear topology from static to dynamic nonlinearities, opening a new route to control topological transport through time‑varying interactions and laying a foundation for non‑equilibrium nonlinear topological matter.
	
	These results are relevant to ongoing experiments in synthetic quantum matter, from dynamically modulated photonic waveguides and fibers~\cite{Mark2022}, driven-dissipative exciton‑polariton systems~\cite{Carusotto2013,Schneider2017,Sieberer2025}, to thermal atoms with tunable nonlinearity and dissipation. By establishing non‑equilibrium dynamics as a practical control dimension, our findings point toward novel forms of dynamically configurable nonlinear topological effects.
	
	\section*{Acknowledgments}
	We thank Qidong Fu, Ming Gong, Nan Li, Fangwei Ye, Biao Wu, Wei Yi, Qi Zhang, Wei Zheng for stimulating discussions and useful help. This work was supported by the Key Project of the National Natural Science Foundation of China Joint Funds (Grants No.~U25A20197), and the National Natural Science Foundation of China (Grants No. 12374246). Z.X.L is supported by the National Natural Science Foundation of China (Grant No. 12574301) and the Zhejiang Provincial Natural Science Foundation of China under Grant No. LZ25A040004. L-C Zhao is supported by the National Natural Science Foundation of China (Contracts No. 12375005, No. 12235007 and No. 12247103). Y.H. acknowledges support by Beijing National Laboratory for Condensed Matter Physics (Grant No. 2023BNLCMPKF001).
	
	\bibliography{Reference}

\end{document}


\title{Supplementary Material for ``Dissipation-induced Nonlinear Topological Gear Switching''}
	
	\author{Xuzhen Cao}
	\thanks{These authors contribute equally to this work.}
	\affiliation{State Key Laboratory of Quantum Optics Technologies and Devices, Institute of Laser Spectroscopy, Shanxi University, Taiyuan 030006, China} 
	\affiliation{Collaborative Innovation Center of Extreme Optics, Shanxi University, Taiyuan 030006, China}
	\affiliation{School of Computer and Information Technology, Shanxi University, Taiyuan 030006, China}
	
	\author{Xiaolin Li}
	\thanks{These authors contribute equally to this work.}
	\affiliation{School of Physics, Northwest University, Xi'an 710127, China}
	\affiliation{Graduate School of China Academy of Engineering Physics, Beijing 100193, China}
	
	\author{Liang Bai}
	\affiliation{Key Laboratory of Computational Intelligence and Chinese Information Processing of Ministry of Education, Institute of Intelligent Information Processing, Shanxi University, Taiyuan 030006, China}

	\author{Zhaoxin Liang}
	\email{zhxliang@zjnu.edu.cn}
	\affiliation{Department of Physics, Zhejiang Normal University, Jinhua 321004, China}
	
	\author{Li-Chen Zhao}
	\email{zhaolichen3@nwu.edu.cn}
	\affiliation{School of Physics, Northwest University, Xi'an 710127, China}
	\affiliation{Shaanxi Key Laboratory for Theoretical Physics Frontiers, Xi'an 710127, China}
	\affiliation{NSFC-SPTP Peng Huanwu Center for Fundamental Theory, Xi'an 710127 and Hefei 230026, China}
	
	\author{Ying Hu}
	\email{huying@sxu.edu.cn}
	\affiliation{State Key Laboratory of Quantum Optics Technologies and Devices, Institute of Laser Spectroscopy, Shanxi University, Taiyuan 030006, China} 
	\affiliation{Collaborative Innovation Center of Extreme Optics, Shanxi University, Taiyuan 030006, China}
	
	\maketitle
	
	In the Supplementary Material, we provide more details on numerical methods, characterization of soliton properties, instantaneous stable nonlinear eigenstates, and validation of the effective model, etc. The Supplementary Material is organized as follows. In Sec.~S1, we elaborate on the spatiotemporal evolution scheme based on the fourth-order Runge-Kutta (RK4) algorithm. In Sec.~S2, we present solutions of the instantaneous nonlinear eigenstates in both conservative and dissipative systems, as well as the effective model. In Sec.~S3, we present an additional phase diagram obtained under different conditions than those in the main text. In Sec.~S4, we verify our ansatz, and in Sec.~S5, we discuss different examples showing quantized transport despite the presence of an aperiodic nonlinearity.

\section{S1. Numerical Simulation}\label{Sec:RK4}
In this section, we present details on our numerical simulations of the dimensionless partial differential equation using the fourth-order Runge-Kutta (RK4) scheme. We also detail our characterization of solitons during propagation. 

\subsection{A. Numerical scheme}

As in the main text, the governing equation of the system is as follows:
\begin{align}\label{Eq:CGLE}
	i\frac{\partial \psi}{\partial z}=\left[-\frac{1}{2}\frac{\partial^2}{\partial \tau^2}+V_{\text{p}}(\tau,z)-g|\psi|^2\right]\psi+i\left[\beta\frac{\partial ^2}{\partial \tau^2}+\delta+\epsilon |\psi|^2\right]\psi.
\end{align}
where $\psi(\tau,z)$ is the complex field, $V_{\text{}}(\tau,z)$ denotes the external potential, $g$ is the nonlinearity coefficient, and $\beta$, $\delta$, $\epsilon$ are dissipative parameters. 

We employ a standard RK4 method to advance the solution in the $z$-direction. Denote the step size by $\Delta z$. The evolution is discretized over $N_{\text{step}} = z_f / \Delta z$ steps, with $z_f$ the total propagation length. At the $n$-the step, the wavefunction is $\psi_n(\tau) \equiv \psi(\tau, z_n)$ with $z_n = n\Delta z$. The RK4 scheme proceeds via four intermediate stages:

\begin{equation}
	\begin{aligned}
		k^{(1)} &= \left(\beta + \frac{i}{2}\right)\frac{\partial^2 \psi_n}{\partial \tau^2} 
		+ \left[\delta - i V_{\text{p}}(\tau, z_n)\right]\psi_n 
		+ (\epsilon + i)|\psi_n|^2 \psi_n, \\
		\psi^{(1)} &= \psi_n + \frac{\Delta z}{2} k^{(1)}, \\
		k^{(2)} &= \left(\beta + \frac{i}{2}\right)\frac{\partial^2 \psi^{(1)}}{\partial \tau^2} 
		+ \left[\delta - i V_{\text{p}}(\tau, z_n + \tfrac{\Delta z}{2})\right]\psi^{(1)} 
		+ (\epsilon + i)|\psi^{(1)}|^2 \psi^{(1)}, \\
		\psi^{(2)} &= \psi_n + \frac{\Delta z}{2} k^{(2)}, \\
		k^{(3)} &= \left(\beta + \frac{i}{2}\right)\frac{\partial^2 \psi^{(2)}}{\partial \tau^2} 
		+ \left[\delta - i V_{\text{p}}(\tau, z_n + \tfrac{\Delta z}{2})\right]\psi^{(2)} 
		+ (\epsilon + i)|\psi^{(2)}|^2 \psi^{(2)}, \\
		\psi^{(3)} &= \psi_n + \Delta z\, k^{(3)}, \\
		k^{(4)} &= \left(\beta + \frac{i}{2}\right)\frac{\partial^2 \psi^{(3)} }{\partial \tau^2} 
		+ \left[\delta - i V_{\text{p}}(\tau, z_n + \Delta z)\right]\psi^{(3)} 
		+ (\epsilon + i)|\psi^{(3)} |^2 \psi^{(3)} .
	\end{aligned}
\end{equation}
The wavefunction at the next propagation step is then given by
\begin{align}
	\psi_{n+1} = \psi_n + \frac{\Delta z}{6}\left(k^{(1)} + 2k^{(2)} + 2k^{(3)} + k^{(4)}\right).
	\label{eq:RK4_update}
\end{align}
In the simulations, the computational domain is chosen to be much larger than the characteristic width of the dissipative solitons to minimize boundary effects.

This approach ensures high accuracy and stability for the numerical simulation of dissipative soliton dynamics, making it suitable for studying complex nonlinear phenomena under various dissipative conditions. To validate our implementation, we benchmark against MATLAB’s adaptive-step \texttt{ode45} solver.

\begin{figure*}[tp]
	\centering
	\includegraphics[width=0.25\textwidth]{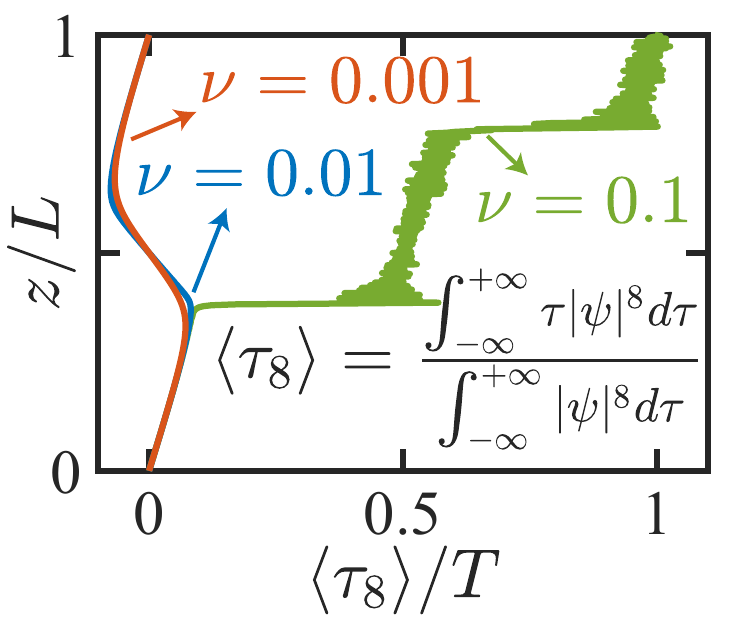} 
	\caption{Higher-order center-of-mass $\langle \tau_8\rangle$ corresponding to Figs.~1(d)-(f) of the main text. Results are shown, respectively, for slow pump speeds $\nu = 0.001$ (red), $\nu = 0.01$ (blue), and $\nu = 0.1$ (green).}
	\label{Fig:FigSup1}
\end{figure*}

\subsection{B. Higher-order center-of-mass displacement and participation ratio}

Here, we take the example of the soliton dynamics shown in Fig.~1 of the main text and provide additional characterization of the soliton properties. 

The crucial quantity characterizing the motion of soliton is its center-of-mass. Typically, the standard quantity $\langle\tau\rangle= {\int_{-\infty}^{+\infty} \tau |\psi|^2 \, d\tau}/{\int_{-\infty}^{+\infty} |\psi|^2 \, d\tau}$ provides a good description, provided the nonlinear energy band is smooth, so that adiabatic condition is well satisfied. If the nonlinear energy band develops discontinuity, such as the nonlinearly induced loop structure and hence a breakdown of strict adiabaticity, the soliton can experience radiation. In this case, $\langle\tau\rangle$ may not be suitable due to the radiation into the dispersive modes. To isolate the displacement of the soliton itself, we follow Ref.~\cite{Jurgensen2023} and quantitatively characterize the displacement of the soliton's center-of-mass through a high-order quantity
\begin{equation}\label{tau8}
	\langle\tau_8\rangle = \frac{\int_{-\infty}^{+\infty} \tau |\psi|^8 \, d\tau}{\int_{-\infty}^{+\infty} |\psi|^8 \, d\tau},
\end{equation}
where $\psi$ is the solution of Eq.~(\ref{Eq:CGLE}). This high-order quantity allows to suppress contributions from the solitonic radiation. In the ideally adiabatic case, both $\langle\tau_8\rangle$ and $\langle\tau\rangle$ quantize after one cycle, while in the nonlinearly induced nonadiabatic case, $\langle\tau_8\rangle$ presents a good quantity. The results of $\langle\tau_8(z)\rangle$ corresponding to Fig.~1 of the text are shown in Fig.~\ref{Fig:FigSup1}, as further confirmation of the topological gear switch.

In addition, to characterize the localization properties of the soliton during propagation, we monitor its peak intensity and the participation ratio (PR) defined by
	\begin{equation}\label{PR}
	\mathrm{PR} = \frac{1}{N_T} \frac{\left( \int_{-N_T/2}^{+N_T/2} |\psi|^2 \, d\tau \right)^2}{\int_{-N_T/2}^{+N_T/2} |\psi|^4 \, d\tau},
\end{equation}
where $N_T$ denotes the system size. The PR quantifies localization: $\mathrm{PR} \to 0$ for strongly localized states and $\mathrm{PR} \to 1$ for extended, delocalized waves. 

\begin{figure*}[tp]
	\centering
	\includegraphics[width=0.5\textwidth]{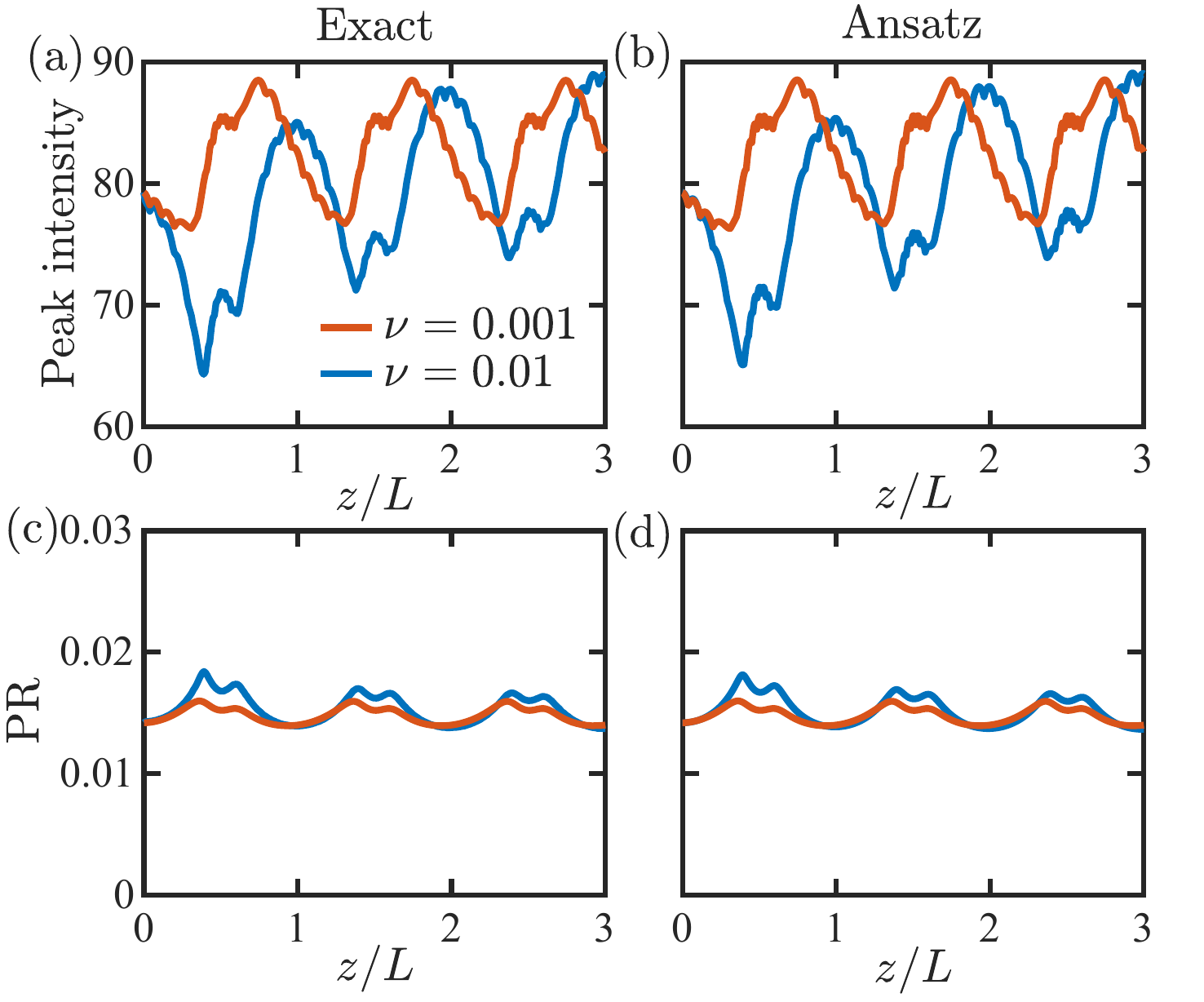}
	\caption{(a), (b) Peak intensity and (c) and (d) participation ratio (PR) of a soliton in a non-Hermitian nonlinear Thouless pump for the slow pump speeds $\nu=0.001,0.1$. (a, c) show exact numerical simulation of Eq.~(\ref{Eq:CGLE}) (with $\delta = 9 \times 10^{-4}$, $\beta = 10^{-5}$, $\epsilon = -10^{-5}$, $g=1$ and $V_s=V_l=25$, initialized in the eigenstate of $H_\textrm{NL}$ with $P(0)=15.2$. (b, d) show corresponding results obtained from the ansatz (\ref{ansatz}). }
	\label{Fig:FigSup2}
\end{figure*}

Figures~\ref{Fig:FigSup2}(a) and (c) show evolutions of the peak intensity and the value of PR during the pumping dynamics in Fig.~1 (d) and (e) of the main text, obtained from numerical solutions of Eq.~(\ref{Eq:CGLE}). We see that the soliton remains well localized during the propagation.  

\section{S2. Instantaneous nonlinear eigenstate}\label{Sec:NSofCS}

In this section, we present numerical details on solutions of the instantaneous stable, localized nonlinear solutions of eigen-equation at each value of $z$ associated with a (i) nonlinear Hermitian Hamiltonian and (ii) nonlinear non-Hermitian operator.

\begin{figure*}[htbp]
	\centering
	\includegraphics[width=0.5\textwidth]{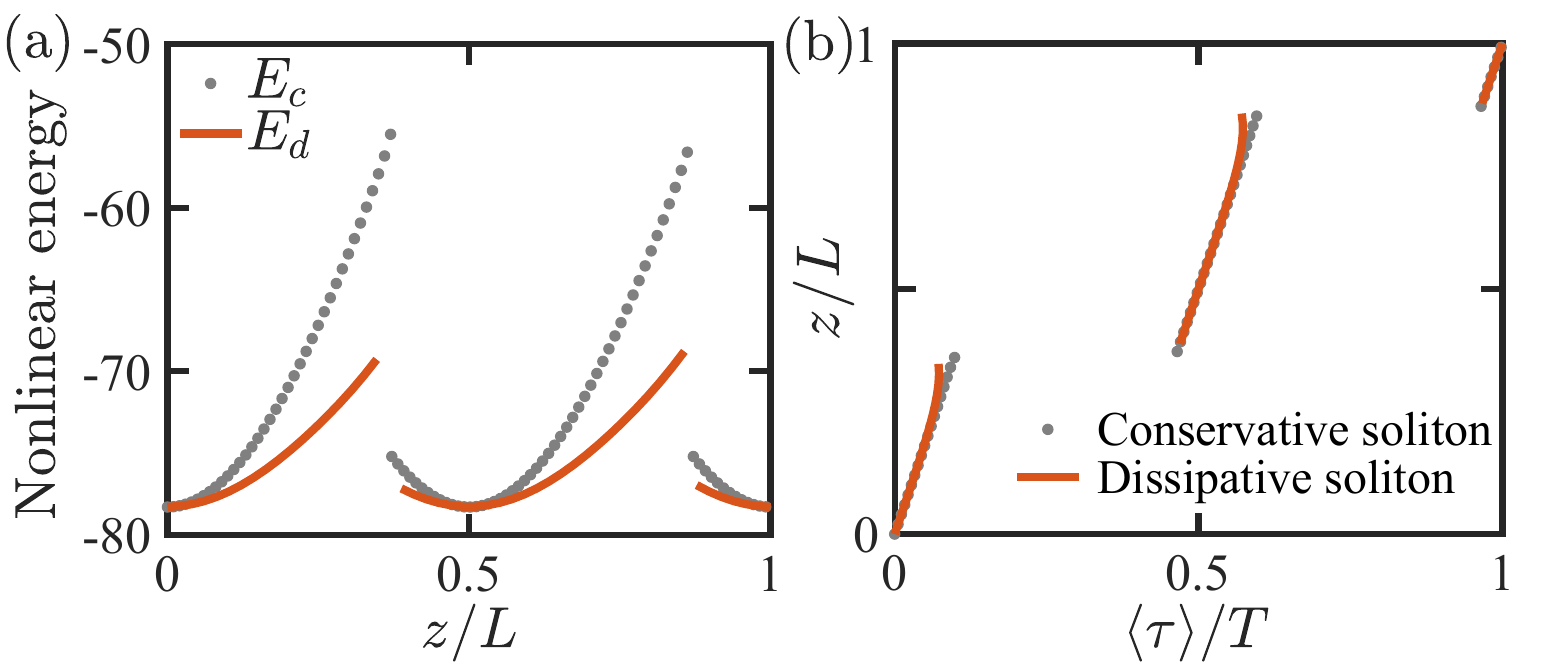}
	\caption{(a) Eigenvalue and (b) the center-of-mass trajectory for the instantaneous stable localized nonlinear eigenstates in conservative and dissipative regimes. Results in the conservative regime correspond to solutions of Eq.~(\ref{Eq:conserH_EEq}) with eigenvalue $E_c$. In the weakly dissipative regime, the instantaneous stable dissipative soliton solutions satisfy Eq.~(\ref{Eq:DissiH_EEq}) with real eigenvalue $E_d$. In (a) and (b), we use $P(0)=15.2$, $\delta = 9 \times 10^{-4}$, $\beta = 10^{-5}$, $\epsilon = -10^{-5}$, and $V_s=V_l=25$.}\label{Fig:FigSup3}
\end{figure*}

\subsection{A. Conservative case}\label{con}

In a conservative nonlinear system, the evolution equation reduces to
\begin{align}\label{Eq:ConservativeCGLE}
	i\frac{\partial \psi}{\partial z} = \left[-\frac{1}{2}\frac{\partial^2}{\partial \tau^2} + V_{\mathrm{p}}(\tau,z) - g|\psi|^2\right] \psi.
\end{align}
To find an instantaneous eigenstate at a fixed $z$, we adopt the ansatz $\psi(\tau,z) = e^{-iE_c z} \varphi_c(\tau)$, which leads to the nonlinear eigenvalue equation 
\begin{align}\label{Eq:conserH_EEq}
	\left[H_0(z)-g|\varphi_c|^2\right]\varphi_c=E_c \varphi_c,
\end{align}
where $H_0=-\frac{1}{2}\frac{\partial^2}{\partial \tau^2} + V_{\mathrm{p}}(\tau,z)$. This equation is solved self-consistently for the mode profile $\varphi_c(\tau)$ and its corresponding eigenvalue $E_c$.

For numerical implementation, we discretize the temporal domain with step size $d\tau$. The second-order derivative in $H_0$ is approximated using the central finite-difference scheme, while the nonlinear term $g|\varphi_c|^2$ becomes a diagonal matrix that depends on the current $\varphi_c$. The full nonlinear eigenvalue problem is solved via a self-consistent iteration scheme~\cite{Tuloup2023}. The implementation proceeds as follows:

\begin{enumerate}
	
	\item Initialize with a Gaussian function $\varphi_{\mathrm{input}}(\tau)$ of norm $P(0)=\int_{-\infty}^{+\infty}|\varphi_{\mathrm{input}}(\tau)|^2d\tau$, with fixed $g=1$. 
	
	\item Construct the nonlinear Hamiltonian $H_{\mathrm{NL}} = H_0 - g|\varphi_{\mathrm{input}}|^2$ using the current input state, and compute all eigenstates $\{|\varphi_n\rangle\}$ of $H_{\mathrm{NL}}$.
	
	\item Select the eigenstate with the largest overlap with $\varphi_{\text{input}}$ as the new iterate $\varphi_{\text{new}}$, and renormalize it to preserve the norm $P(0)$.
	
	\item Repeat steps 2–3 until convergence. The converged state is the instantaneous nonlinear eigenstate at the current $z$, and we also obtain its eigenvalue $E_c$ satisfying Eq.~\eqref{Eq:conserH_EEq}.
	
	\item Use the converged eigenstate as the initial guess for the next $z$-slice ($z + dz$), accelerating convergence.
\end{enumerate}

This process yields an instantaneous nonlinear eigenvalue spectrum as a function of $z$. Furthermore, the nonlinear eigenstate obtained by this method at $z = 0$ is used as the initial state for Figs.~1 and 2 in the main text. The gray point in Fig.~\ref{Fig:FigSup3}(a) shows the instantaneous eigenvalue $E_c$ for norm $P(0) = 15.2$ in the conservative system, which matches the norm of the initial state in Fig.~1 of the main text. The cente-of-mass position of the corresponding eigenstate is shown as the gray point in Fig.~\ref{Fig:FigSup3}(b).

\subsection{B. Weakly dissipative case}\label{dis}

For the dissipative nonlinear system governed by Eq.~\eqref{Eq:CGLE}, the relevant instantaneous nonlinear eigen-equation takes the form
\begin{align}\label{Eq:DissiH_EEq}
	\mathcal{H}_{\mathrm{NH}}\varphi_d=E_d \varphi_d;\hspace{2mm}\mathcal{H}_{\mathrm{NH}} \equiv H_0(z) - g|\varphi_d|^2 + i\left[ \beta\frac{\partial^2}{\partial \tau^2} + \delta + \epsilon |\varphi_d|^2 \right], 
\end{align}
which is governed by a non-Hermitian nonlinear operator $\mathcal{H}_{\mathrm{NH}}$. Our goal is to find solutions of Eq.~(\ref{Eq:DissiH_EEq}) with real eigenvalue $E_d\in \mathbb{R}$ at each $z$, which corresponds to the instantaneous dissipative soliton (DS) stablized by double balances (i.e., instantaneous steady state). 

To this end, we adopt a dynamical approach that allows us to find the steady state at each value of $z$. Specifically, for a given $z$, we treat $V_{\mathrm{p}}(\tau,z)$ as a static potential and compute $\psi_d(\tau; z)$ by evolving Eq.~\eqref{Eq:CGLE} in an auxiliary propagation variable $z'$:
\begin{align}\label{Eq:CGLE_Aux}
	i\frac{\partial \psi}{\partial z'} = \left[ -\frac{1}{2}\frac{\partial^2}{\partial \tau^2} + V_{\text{p}}(\tau,z) - g|\psi|^2 \right]\psi 
	+ i\left[ \beta\frac{\partial^2}{\partial \tau^2} + \delta + \epsilon |\psi|^2 \right]\psi. 
\end{align}
The evolution starts from a Gaussian ansatz informed by the known free-space dissipative soliton solution~\cite{Mao2025}, $\psi(\tau,0) = \sqrt{\frac{3\delta}{\beta - 2\epsilon}} \exp\!\left(-\tau^2/l^2\right)$, with width $l = T/4$. We numerically integrate Eq.~\eqref{Eq:CGLE_Aux} forward in $z'$ until the norm $P(z') = \int_{-\infty}^{+\infty} |\psi(\tau,z')|^2 d\tau$ converges to a steady value, i.e.,  when the imaginary part of the average energy $\int_{-\infty}^{+\infty} \mathcal{H}_{\mathrm{NH}}|\psi(\tau,z')|^2 d\tau\sim 10^{-6}$, much smaller than the system parameters $(\delta, \beta, \epsilon P(0))$. This stead state is take as the stable dissipative soliton attractor $\psi_d(\tau; z)$ associated with the potential configuration at $z$. 

In Fig.~\ref{Fig:FigSup3} (see red curves in (a) and (b)), we show the energy and the center-of-mass position of the instantaneous steady soliton state when $\delta = 9 \times 10^{-4}$, $\beta = 10^{-5}$, and $\epsilon = -10^{-5}$ (see red curves). We see that the instantaneous DS exhibits qualitatively similar behavior as the instantaneous conservation soliton. This is expected, because the dissipative parameters are significantly small compared to the energy scales associated with the coherent process. 

\section{S3. Supplementary phase diagram}\label{Sec:Phase2}

\begin{figure*}[tp]
	\centering
	\includegraphics[width=0.3\textwidth]{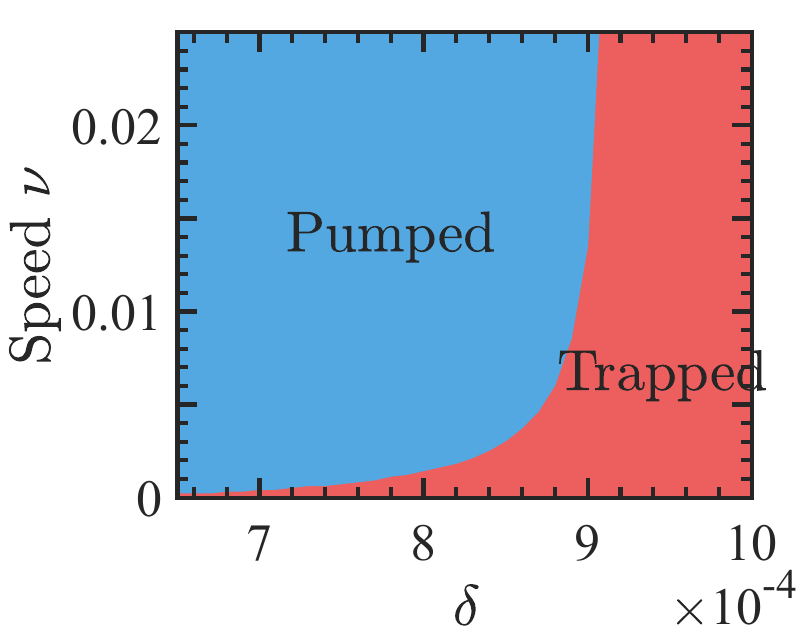}
	\caption{Displacement over one pumping cycle is shown as a function of linear gain $\delta$ and pump speed $\nu$. For other parameters, $\beta = 10^{-5}$, $\epsilon = -10^{-5}$, and $V_s = V_l = 25$. The system is initialized in the instantaneous stable dissipative soliton at $z=0$. }\label{Fig:FigSup4}
\end{figure*}

\begin{figure*}[htbp]
	\centering
	\includegraphics[width=0.5\textwidth]{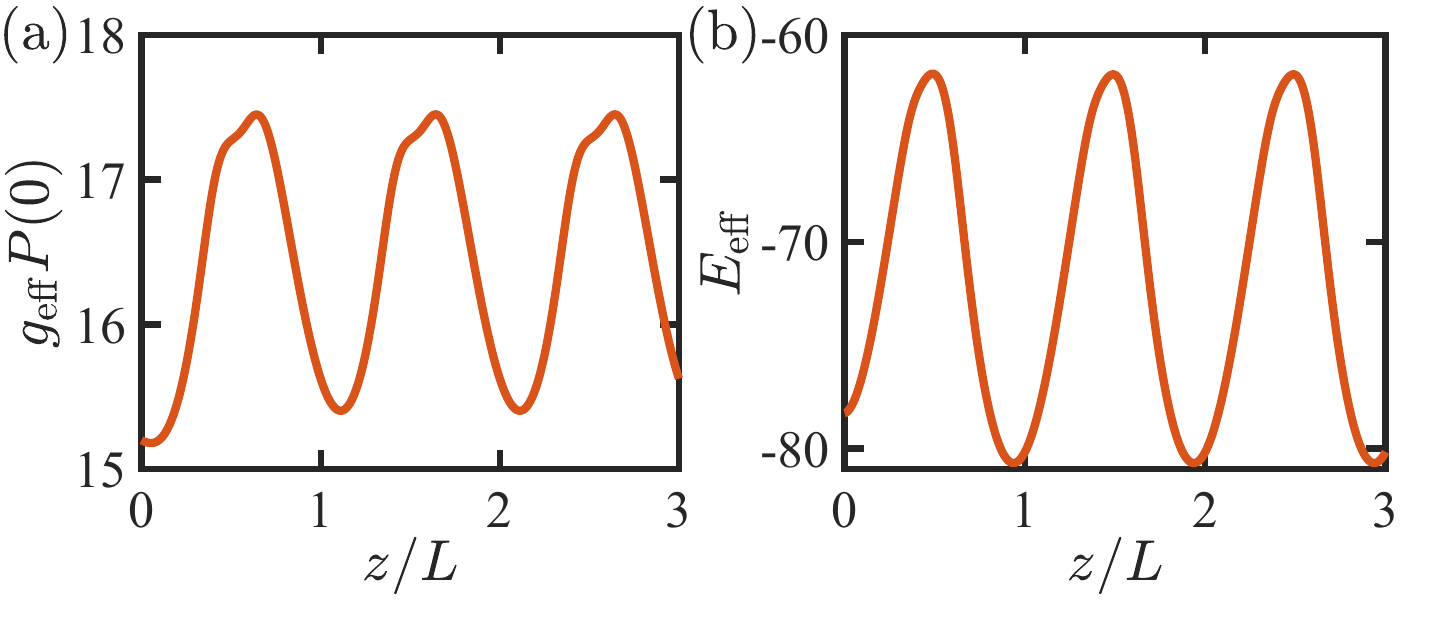}
	\caption{(a) Effective power $g_{\mathrm{eff}}P(0)$ and (b) the effective nonlinear spectrum $E_{\mathrm{eff}}$ over three pump cycles, when $P(0)=15.2$, $\nu=0.001$, $\delta = 9 \times 10^{-4}$, $\beta = 10^{-5}$, $\epsilon = -10^{-5}$, and $V_s=V_l=25$.}
	\label{Fig:FigSup5}
\end{figure*}

In Fig.~2 of the main text, we have demonstrated the speed-dependent topological gear switch, by showing the solitonic displacement after the first cycle as a function of the initial power $P(0)$ and pump speed $\nu$, using system parameters in Fig.~1 of the text. Here, we demonstrate the generality of our key results by varying the linear gain $\delta$. In Fig.~\ref{Fig:FigSup4}, we set $\beta = 10^{-5}$ and $\epsilon = -10^{-5}$ (same as Fig.~2 the main text), and plot the high-order center-of-mass displacement $\langle \tau_8\rangle$ after one period as a function of the linear gain $\delta$ and pump speed $\nu$. For the initial condition, we use the stable instantaneous dissipative soliton at $z=0$ obtained via the approach in Sec.~S2. The numerical results are obtained via solutions of Eq.~\eqref{Eq:CGLE}. Again, the speed-dependent topological phase transition is observed.

\begin{figure*}[tp]
	\centering
	\includegraphics[width=0.9\textwidth]{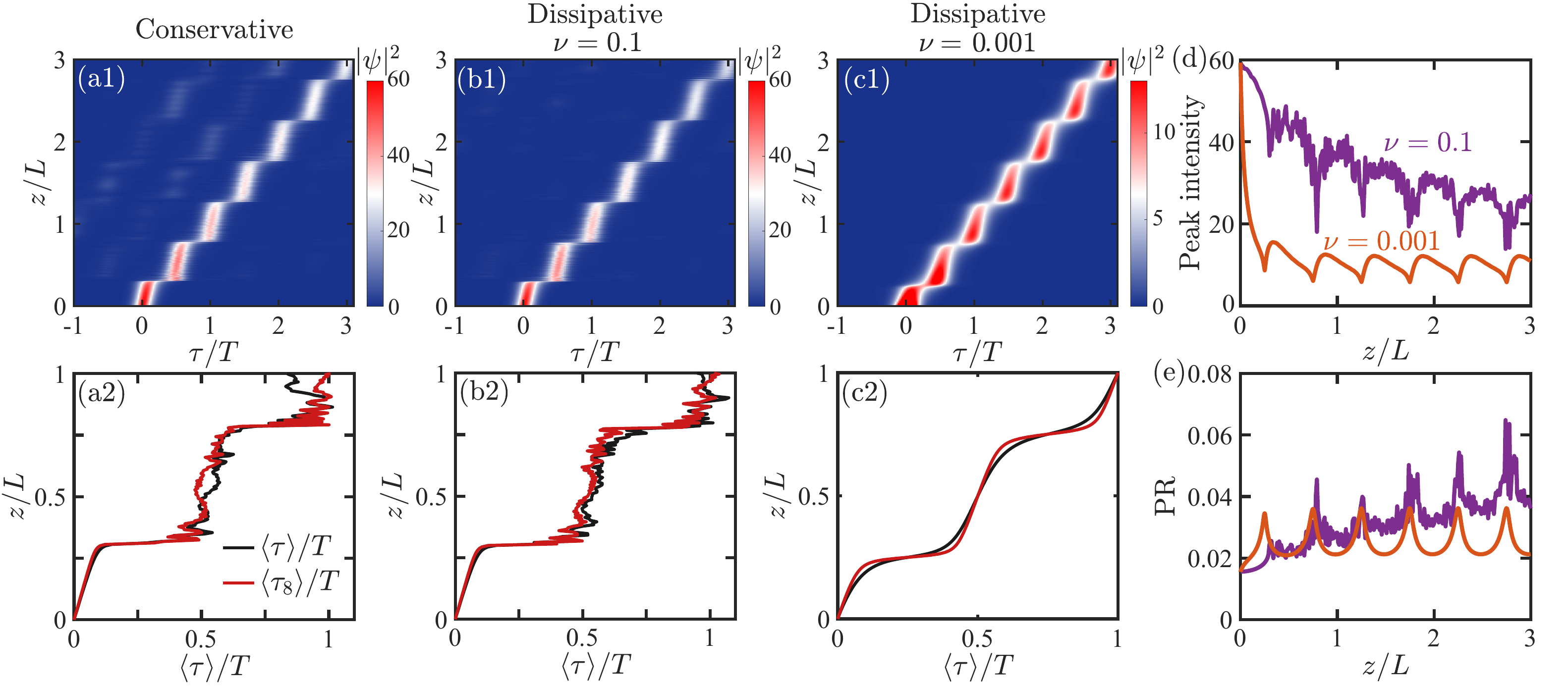}
	\caption{Soliton dynamics corresponding to case I (corresponding to Figs.~4(a) and (b) in the main text). (a1-c1) show propagation of same input soliton in (a1) the conservative regime at $\nu=0.1$ and in the weakly dissipative regimes at (b1) $\nu=0.1$ and (c1) $\nu=0.001$. (a2-c2) show corresponding displacement $\langle\tau\rangle$ and $\langle\tau_8\rangle$ of the center-of-mass of the wavefunction $\psi(z)$. In weakly dissipative regime, we also show (d) peak intensity and (e) PR of the propagating soliton at different pump speed. In all computations, the system is initialized in the instantaneous eigenstate of $H_\textrm{NL}(0)$ with $g=1$, $V_s=V_l=25$, and $P(0)=12.42$. For dissipative parameters, we take $\delta = 10^{-3}$, $\beta = 5\times10^{-5}$, $\epsilon = -5\times10^{-5}$. }
	\label{Fig:FigSup6}
\end{figure*}

\section{S4. Validation of ansatz}\label{Sec:ansatz}

Here we provide validation of the ansatz in Eq.~(3) in the main text. Additionally, we show long-time behaviors of the effective power $g_{\mathrm{eff}}P(0)$ and the effective instantaneous nonlinear eigenvalues. 

As described in the main text, the ansatz is
\begin{equation}\label{ansatz}
	\psi(z,\tau) \approx f(z)\, \varphi_{\mathrm{eff}}(z,\tau), \quad \text{with} \quad f(0) = 1,
\end{equation}
where $\varphi_{\mathrm{eff}}(z,\tau)$ is the solution of the effective nonlinear Schr\"odinger equation:
\begin{equation}\label{effective}	
	\left[ H_0(\tilde{z}) - g_{\mathrm{eff}}(\tilde{z}) |\varphi_{\mathrm{eff}}|^2 \right] \varphi_{\mathrm{eff}} 
	= E_{\mathrm{eff}}(\tilde{z}) \varphi_{\mathrm{eff}},
\end{equation}
where $g_{\mathrm{eff}}(\tilde{z}) = g |f(\tilde{z})|^2$, while $f(\tilde{z})$ evolves according to
\begin{equation}\label{f}
	\frac{\partial f}{\partial \tilde{z}} = \pi \frac{\gamma(\tilde{z})}{\nu} f, \quad 
	\gamma(\tilde{z}) \equiv \frac{\langle \varphi_{\mathrm{eff}}(\tilde{z}) | \mathcal{D} | \varphi_{\mathrm{eff}}(\tilde{z}) \rangle}{P(0)}.
\end{equation}
To validate Eq.~(\ref{ansatz}), in Fig.~\ref{Fig:FigSup2}(b) and (d), we calculate the peak intensity [Fig.~\ref{Fig:FigSup2}(b)] and PR [Fig.~\ref{Fig:FigSup2}(d)] of solitons based on Eq.~(\ref{ansatz}), respectively, and compare them with those obtained from solutions of Eq.~\eqref{Eq:CGLE} [Fig.~\ref{Fig:FigSup2}(a) and (c)]. A good agreement with each other is found. 

In addition, we note that Figs.~3(b) and (c) of the main text only presented the effective instantaneous nonlinear eigenstates in one pump cycle. Here, we present the effective nonlinearity $g_{\mathrm{eff}} P(0)$ [Fig.~\ref{Fig:FigSup5}(a)] and instantaneous nonlinear eigenvalue $E_{\mathrm{eff}}$ [Fig.~\ref{Fig:FigSup5}(b)] over three pump cycles for $\nu = 0.001$. We see that while the aperiodicity in $g_{\mathrm{eff}}(z)P(0)$ is prominent at the first cycle, it diminishes over successive pump cycles. For longer times, both the effective nonlinearity $g_{\mathrm{eff}}(z)P(0)$ and $E_{\mathrm{eff}}$ seem to approach a periodic oscillation, indicating that the soliton converges to a limit cycle near the attractor governed by gain-loss balance in the long-time limit.

\section{S5. Quantization despite aperiodic nonlinear pumping}\label{Sec:QuantizedPumping}

In this section, we discuss three different pumping scenarios with aperiodic effective nonlinearity. In all computations, our initial conditions are the instantaneous nonlinear eigenstates obtained through self-consistent iteration at $z = 0$ in a corresponding conservative system ($\delta = \beta = \epsilon = 0$).

\subsection*{Case I: Supplementary results to Fig.~4 of main text}

We begin with the case corresponding to Fig.~4 (a) in the main text, supplementing addition information. Specifically, we analyze propagation of the same input soliton in the conservative regime and weakly dissipative regime ($\delta = 10^{-3}$, $\beta = 5 \times 10^{-5}$, and $\epsilon = -5 \times 10^{-5}$) at different pump speed $\nu$, as shown in Fig.~\ref{Fig:FigSup6}.

\begin{figure*}[t]
	\centering
	\includegraphics[width=0.5\textwidth]{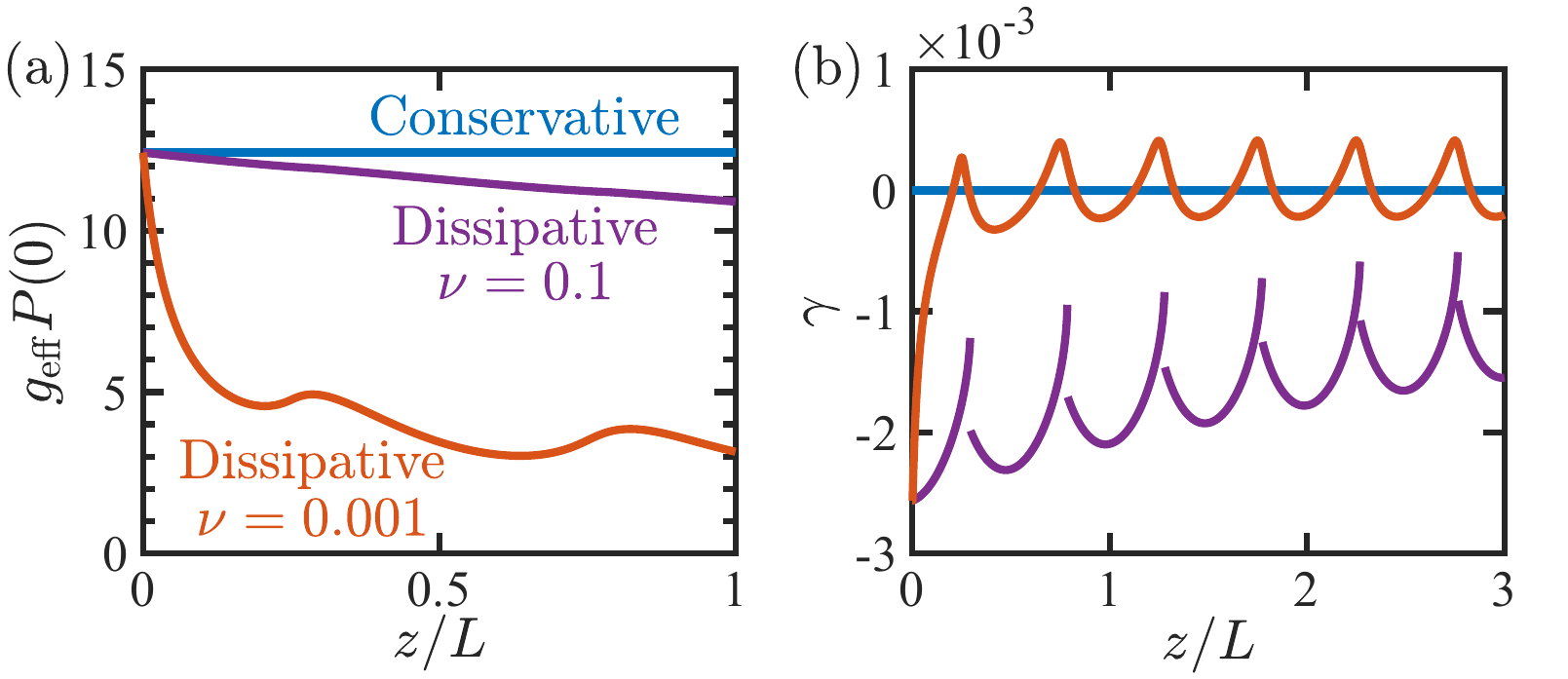}
	\caption{Effective model corresponding to Fig.~\ref{Fig:FigSup6}. 
		(a) Effective power $g_{\mathrm{eff}}P(0)$ and (b) effective dissipation rate $\gamma$ [Eq.~(\ref{f})] at $\nu = 0.1, 0.001$. Results for conservative case are shown as a reference point.}
	\label{Fig:FigSup7}
\end{figure*}

\begin{figure*}[t]
	\centering
	\includegraphics[width=0.6\textwidth]{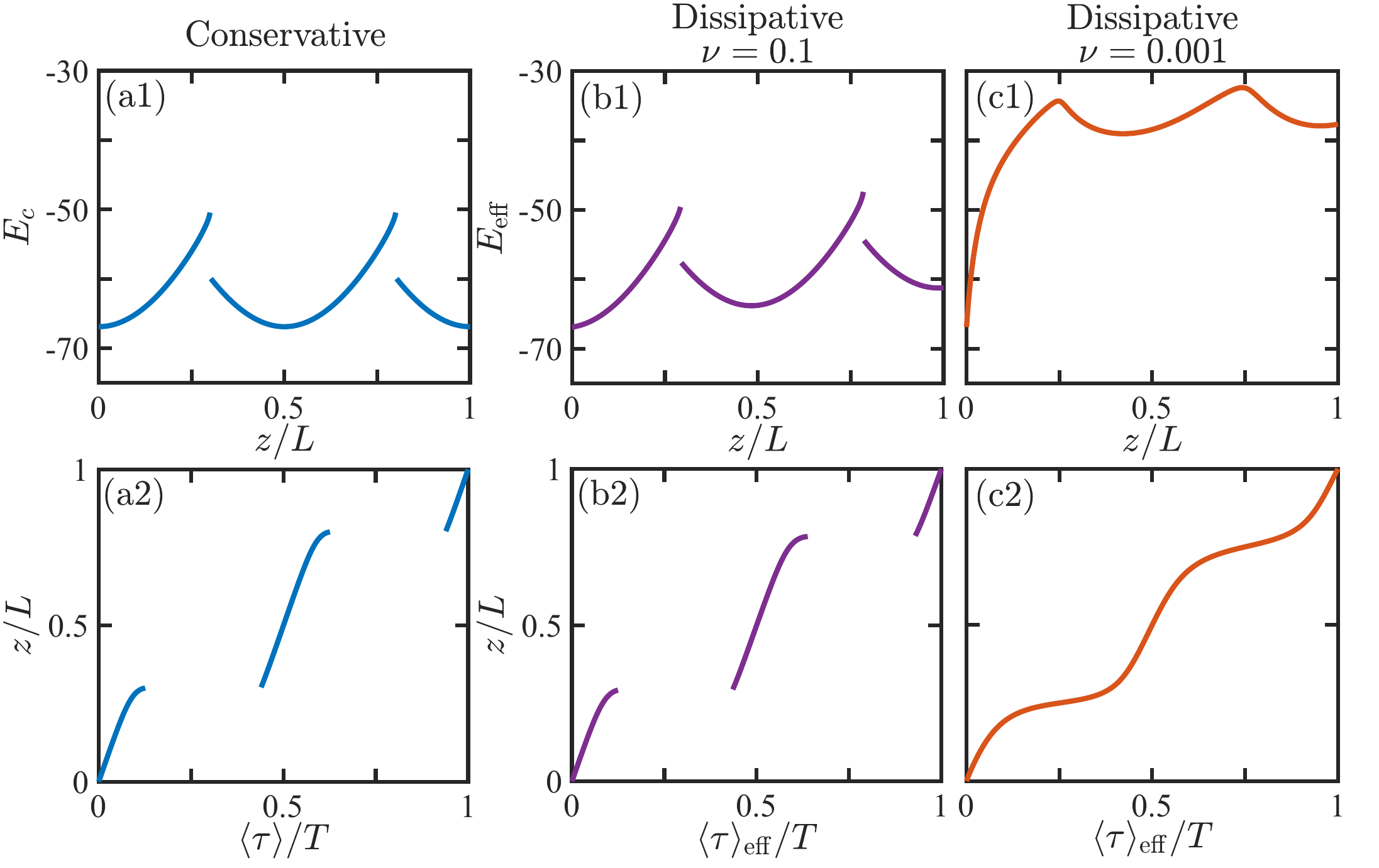}
	\caption{(Effective) instantaneous stable soliton corresponding to Fig.~\ref{Fig:FigSup6} in (a1)-(a2) conservative regime and (b1-c2) weakly dissipative regime when (b1)-(b2) $\nu = 0.1$ and (c1)-(c2) $\nu = 0.001$. (a1)-(c1) show the (effective) instantaneous eigenvalue of the corresponding (effective) instantaneous stable soliton. (a2)-(c2) show the center-of-mass position of the (effective) instantaneous stable soliton.	}
	\label{Fig:FigSup8}
\end{figure*}

In the conservative regime [Fig.~\ref{Fig:FigSup6}(a1)], for the bare nonlinearity $gP(0)=12.42$, the soliton is pumped regardless of speed $\nu$. Fig.~\ref{Fig:FigSup6}(a2) shows that $\langle \tau\rangle$ is not quantized but $\langle \tau_8\rangle$ is quantized. This is because for the nonlinear strength $gP(0)=12.42$, the nonlinear energy band $E_c(z)$ develops discontinuities [Fig.~\ref{Fig:FigSup8}(a1)], which are reminiscent of the familiar loop structure. This characteristic nonlinear feature induces non-adiabatic effect~\cite{Wu2000,Liu2002,Tuloup2023}, causing the soliton to radiate its energy into dispersive modes. Despite this radiation, the center-of-mass of the localized soliton itself undergoes quantized displacement, which is well captured by the behavior of $\langle \tau_8\rangle$. This is in agreement with the center-of-mass position of the instantaneous stable soliton of the nonlinear Hamiltonian at every value of $\phi=\nu z$ [Fig.~\ref{Fig:FigSup8}(a2)].

\begin{figure*}[t]
	\centering
	\includegraphics[width=0.75\textwidth]{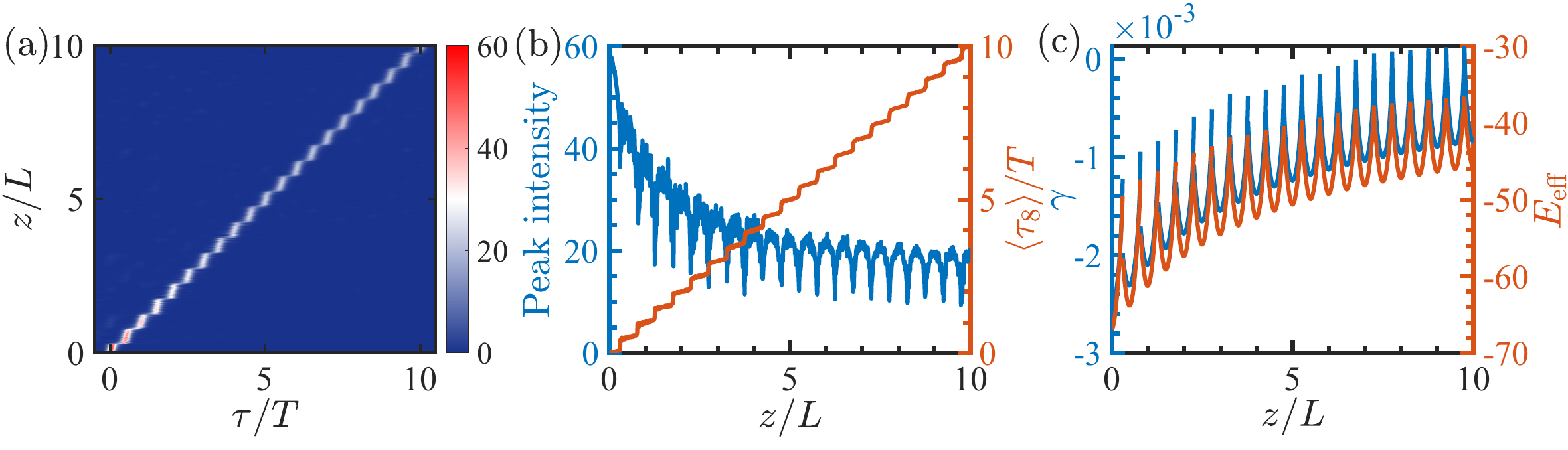}
	\caption{Long-term evolution over ten pumping cycles, corresponding to Fig.~\ref{Fig:FigSup6}(b). Shown are (a) Intensity distribution, (b) Peak intensity and center-of-mass displacement quantified by $\langle\tau_8\rangle$, (c) Effective dissipation $\gamma(z)$ in Eq.~(\ref{f}) and effective instantaneous eigenvalue $E_{\rm{eff}}(z)$. The initial state and all other parameters are identical to those in Fig.~\ref{Fig:FigSup6}(b).}
	\label{Fig:FigSup9}
\end{figure*}

In the weakly dissipative regime and for $\nu=0.1$ [Fig.~\ref{Fig:FigSup6}(b1)], the pumping dynamics closely mimics the conservative counterpart, showing quantized $\langle \tau_8\rangle$ but nonquantized $\langle \tau\rangle$ [Fig.~\ref{Fig:FigSup6}(b2)]. This can be explained through an effective Hermitian Hamiltonian with an effective nonlinearity $g_\textrm{eff}(z)P(0)$, i.e., for $\nu=0.1$ and correspondingly $\gamma/\nu\sim 0.001$, one has $g_\textrm{eff}(z)\approx g$ [Fig.~\ref{Fig:FigSup7}(a)]. As shown in Fig.~\ref{Fig:FigSup8}(b1), the corresponding effective energy spectrum exhibits discontinuities like its conservative counterpart, thus causing non-adiabatic effect manifested in the noisy dynamics in Figs.~\ref{Fig:FigSup8}(d) and (e) (see purple curves). Therefore, not surprisingly, $\langle\tau\rangle$ is not quantized, but $\langle \tau_8\rangle$ remains quantized [Fig.~\ref{Fig:FigSup6}(b2)], in accordance with the trajectory of the instantaneous effective stable soliton [Fig.~\ref{Fig:FigSup8}(b2)]. 

At much slower speed $\nu=0.001$, the soliton is pumped [Fig.~\ref{Fig:FigSup6}(c1)], with both $\langle \tau_8\rangle$ and $\langle \tau\rangle$ are quantized [Fig.~\ref{Fig:FigSup6}(c2)]. Again, this can be understood by looking at the effective model. In this case, the effective nonlinearity drops substantially from the conservative counterpart [Fig.~\ref{Fig:FigSup7}(a)]. The resulting effective energy spectrum is substantially modified and becomes continuous [Fig.~\ref{Fig:FigSup8}(c1)]. This indicates adiabatic condition can be well preserved even in the presence of nonlinearity, as evidenced by the smooth dynamics in Figs.~\ref{Fig:FigSup6}(d) and (e) (see red curves). Therefore, we expect quantization of both $\langle\tau\rangle$ and $\langle\tau_8\rangle$ at the end of a cycle. Indeed, the corresponding trajectory of instantaneous effective stable soliton becomes contiguous, with a displacement by one unit [Fig.~\ref{Fig:FigSup8}(c2)]. 

At long times, the gain and loss are expected to eventually reach a dynamical equilibrium for both $\nu=0.1$ and $\nu=0.001$. To see this, in Fig.~\ref{Fig:FigSup7}(b) we plot the evolution of $\gamma(z)$ according to Eq.~(\ref{f}) for the dissipative parameters in Fig.~1 of the main text. Indeed, $\gamma(z)$ equilibrates into an oscillation around a zero mean value, which is potentially related to the limit cycle around a fixed point. This dynamical equilibrium is also manifested in the solitonic dynamics in Figs.~\ref{Fig:FigSup6}(d) and (e), where we see that although the PR and peak intensity of the soliton changes appreciably over the first cycle, it gradually equilibrate over more pump cycles. Note that the dynamical balance is reached much faster at $\nu=0.001$ than for $\nu=0.1$. For the latter, we observe the dynamical equilibrium builds up over more than ten pump cycles [Fig.~\ref{Fig:FigSup9}].

\subsection*{Case II: Effective nonlinear Hermitian Hamiltonian is slowly quenched from linear to nonlinear regimes}

\begin{figure*}[tp]
	\centering
	\includegraphics[width=0.9\textwidth]{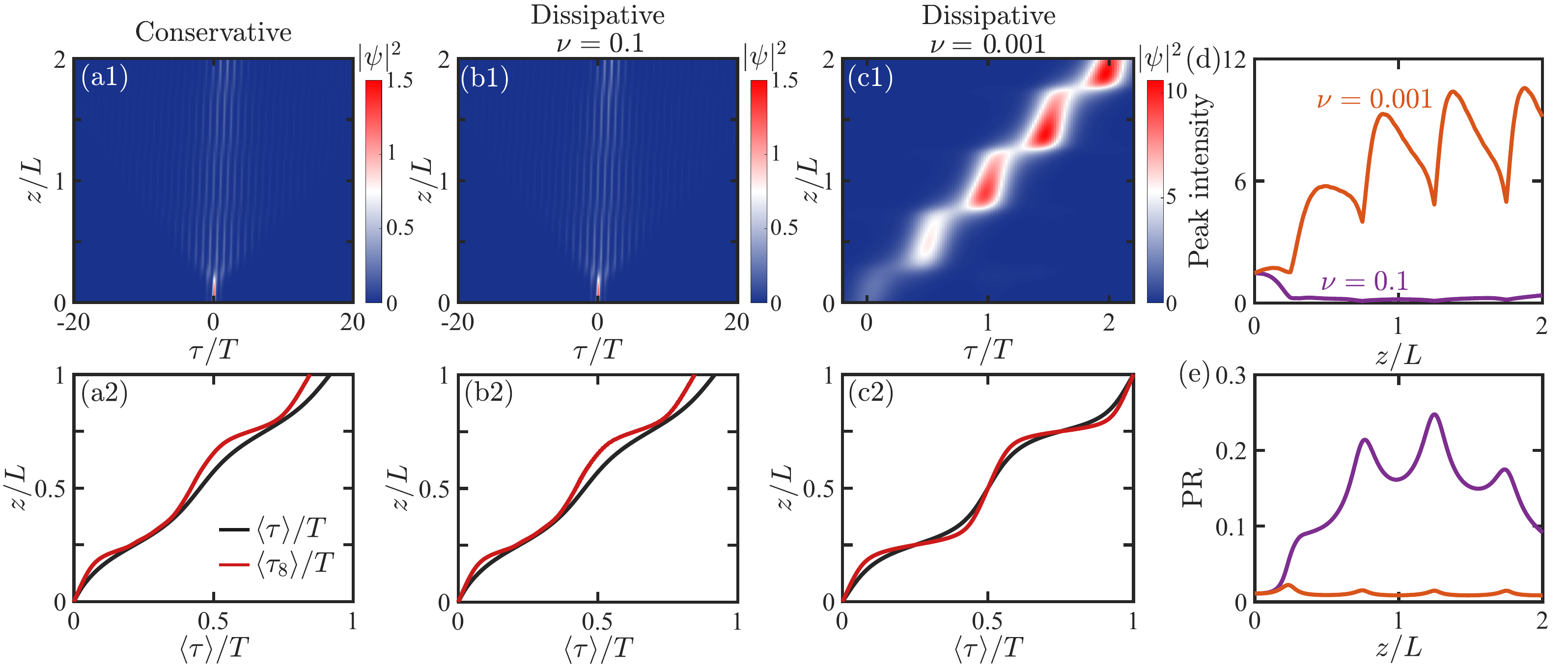}
	\caption{Pumping dynamics in case II (corresponding to Figs.~4(b) in the main text). (a1-c1) show propagation of same input wave function in (a1) the conservative regime at $\nu=0.1$ and in the weakly dissipative regimes at (b1) $\nu=0.1$ and (c1) $\nu=0.001$. (a2-c2) show corresponding displacement $\langle\tau\rangle$ and $\langle\tau_8\rangle$ of the center-of-mass of the wavefunction $\psi(z)$. In weakly dissipative regime, we also show (d) peak intensity and (e) PR of the propagating soliton at different pump speed. In all computations, the system is initialized in the instantaneous eigenstate of $H_\textrm{NL}(0)$ with $g=1$, $V_s=V_l=25$ and $P(0) = 0.5$ (corresponding peak intensity is $|\psi_0|_{\mathrm{max}}^2 \approx 1.47$). For dissipative parameters, we take $\delta = 1.2 \times 10^{-3}$, $\beta = 6.5 \times 10^{-5}$, $\epsilon = -6.5 \times 10^{-5}$.}
	\label{Fig:FigSup10}
\end{figure*}

We now examine a case with the bare nonlinearity $gP(0)=0.5$ (corresponding to Fig.~4 (b) in the main text), where the pumping lies in the linear regime in the absence of dissipation. Indeed, as shown in Fig.~\ref{Fig:FigSup10}(a1), starting from the eigenstate of $H_0(0)-g|\psi|^2$, the wave packet rapidly disperse without forming a soliton. In this linear regime, the center-of-mass of wave function does not show quantized displacement [Fig.~\ref{Fig:FigSup10}(a2)], due to a nonuniform population of the topological Bloch band. 

Under weakly dissipation ($\delta = 1.2 \times 10^{-3}$, $\beta = 6.5 \times 10^{-5}$, $\epsilon = -6.5 \times 10^{-5}$), however, the dynamics become strongly speed-dependent.  At $\nu = 0.1$, the wave packet still rapidly disperses [Fig.~\ref{Fig:FigSup10}(b1)], without soliton formation as supported by the computation of peak intensity and PR [purple curves in Figs.~\ref{Fig:FigSup10}(d) and (e)]. In sharp contrast, at $\nu = 0.001$, a stable localized soliton is formed on the fly [Fig.~\ref{Fig:FigSup10}(c1), red curves in Figs.~\ref{Fig:FigSup10}(d) and (e)], and its center-of-mass shifts by a unit at the end of each cycle [Fig.~\ref{Fig:FigSup10}(c2)]. This quantized transport is remarkable, as the system evolution is highly aperiodic when dynamically transiting from the initial linear to final nonlinear regimes, so that the states at the beginning and end of the cycle are even different in nature.

\begin{figure*}[tp]
	\centering
	\includegraphics[width=0.5\textwidth]{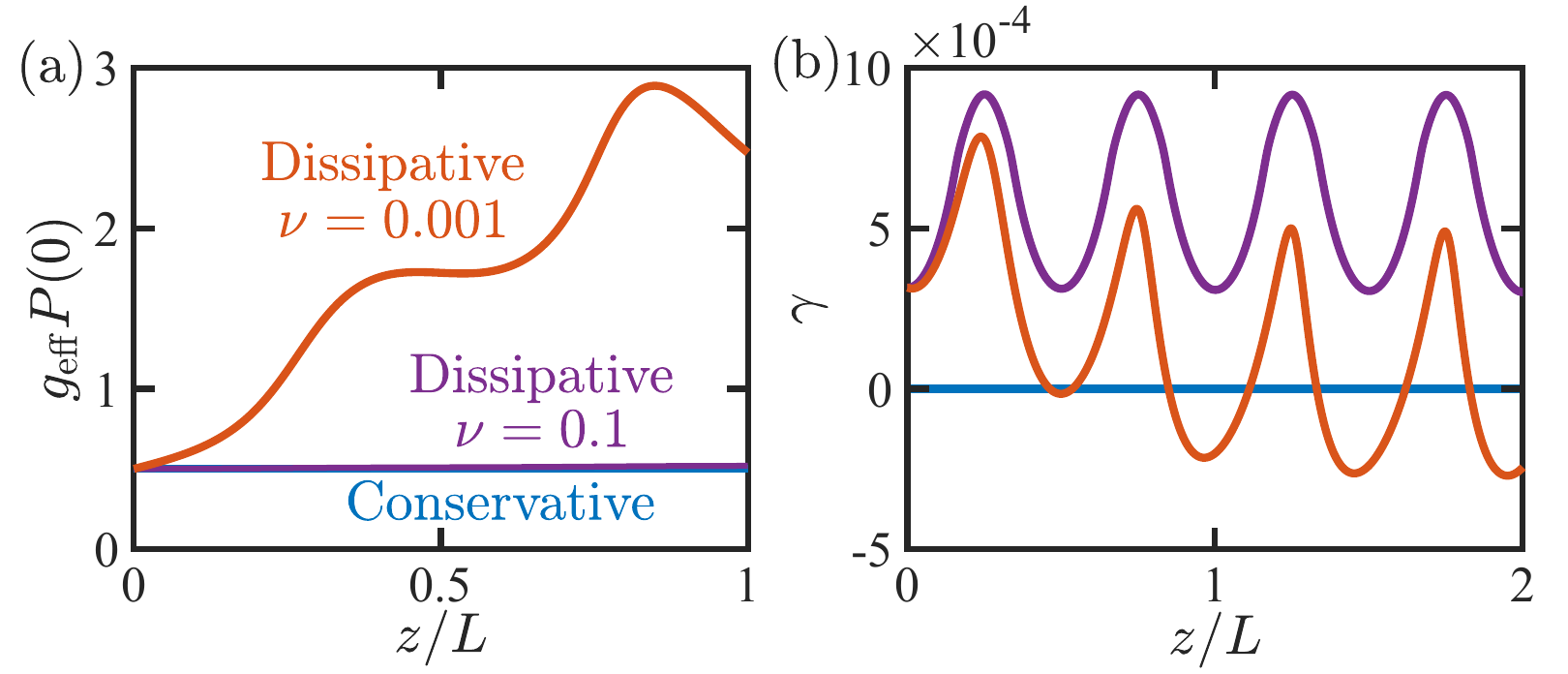}
	\caption{Effective model corresponding to Fig.~\ref{Fig:FigSup10}. 
		(a) Effective power $g_{\mathrm{eff}}P(0)$ and (b) effective dissipation rate $\gamma$ [Eq.~(\ref{f})] at $\nu = 0.1, 0.001$. Results for conservative case are shown as a reference point.}
	\label{Fig:FigSup11}
\end{figure*}

\begin{figure*}[ht]
	\centering
	\includegraphics[width=0.45\textwidth]{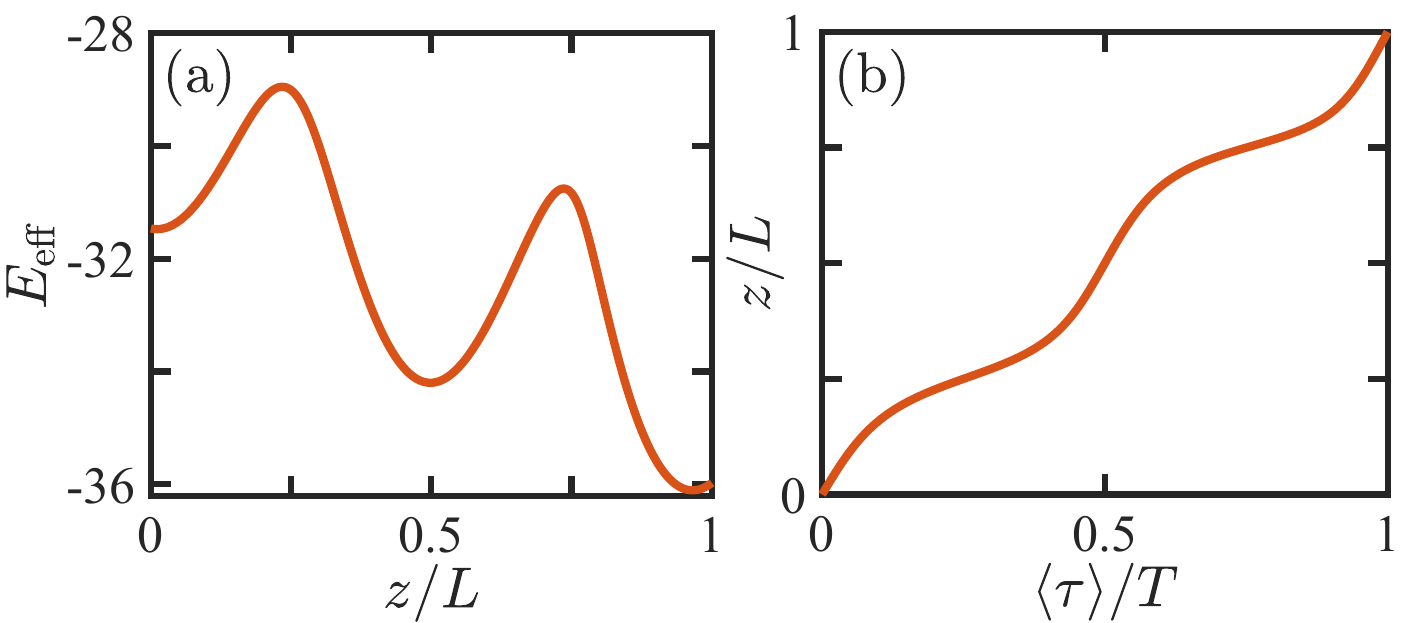}
	\caption{Effective instantaneous stable soliton corresponding to Fig.~\ref{Fig:FigSup10} in weakly dissipative regime when $\nu = 0.001$. (a)-(b) show (a) the effective instantaneous eigenvalue and (b) center-of-mass of the corresponding effective instantaneous stable soliton.}
	\label{Fig:FigSup12}
\end{figure*}

Still, such speed-dependent transition from linear transport to quantized nonlinear transport can be well explained by our effective model. For $\nu=0.1$, the effective nonlinearity $g_{\mathrm{eff}}P(0)$ remains close to the bare value [Fig.~\ref{Fig:FigSup11}(a)], indicating the pumping remains in the linear regime, leading to the dispersion observed in Fig.~\ref{Fig:FigSup10}(b1). However, for $\nu=0.001$, $g_{\mathrm{eff}}P(0)$ is strongly modified to deviate from the conservative case [Fig.~\ref{Fig:FigSup11}(a)], pushing the effective dynamics crossovers into a strongly nonlinear regime where stable soliton can be formed. The resulting effective nonlinear energy spectrum is continuous [Fig.~\ref{Fig:FigSup12}(a)] and suggests adiabatic evolution throughout the pumping process. As such, the input wave packet follows the center-of-mass position of the instantaneous nonlinear eigenstates [Fig.~\ref{Fig:FigSup12}(b)], which forms contiguous trajectory that shifts by unit at the end of each cycle. Note that even in this dramatic case, the system at long times eventually reaches a dynamical equilibration, as indicated by Fig.~\ref{Fig:FigSup11}(b).

\begin{figure*}[t]
	\centering
	\includegraphics[width=0.45\textwidth]{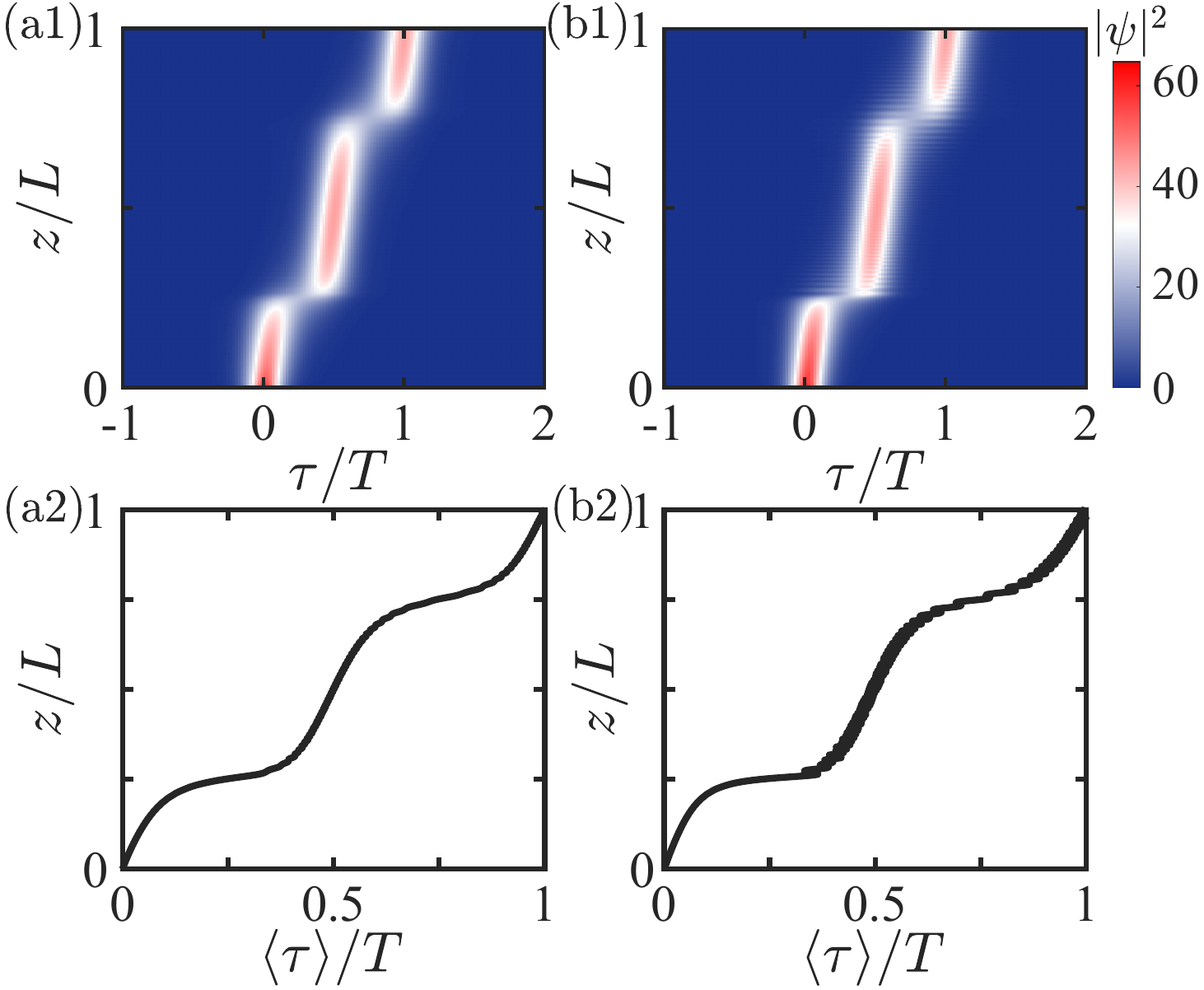}
	\caption{Soliton dynamics in case III, i.e., a purely conservative nonlinear Thouless pumping with aperiodic, time-varying nonlinearity. Results of (a1, b1) intensity distribution and (a2, b2) center-of-mass displacement are obtained by solutions of Eq.~(\ref{Eq:NSE}) when $g(z)$ takes the form (a1,a2) identical to $g_{\mathrm{eff}}$ in Fig.~3(b) with $\nu=0.001$, and (b1,b2) $g(z) = g_0 + (g_f - g_0) (1 - e^{-z/\sigma})$ with $g_0=1$, $g_f=0.2536$ and $\sigma=2\pi$. In both cases, the system is initialized in the same state as in Fig.~3. }
	\label{Fig:FigSup13}
\end{figure*}

\subsection*{Case III: purely conservative pumping with aperiodic time-varying nonlinearity}

So far, we have shown that the physics of nonlinear Thouless pumping in the weakly dissipative regime can be captured by an effective Hermitian model with aperiodic, time-varying nonlinearity. This suggests that even for a purely conservative system with time-varying nonlinearity $g(z)$, it is possible to achieve quantization. Here, we explore further into this aspect by studying the pumping dynamics governed by the following nonlinear Schr$\ddot{o}$dinger equation, 
\begin{equation}\label{Eq:NSE}
	i\frac{\partial \psi}{\partial z}=\left[H_0(z)-g(z)|\psi|^2\right]\psi, 
\end{equation}
where $g(z)$ takes different forms of time-dependence. 

To benchmark our discussion, we begin with an example where $g(z)$ takes the same form as $g_\textrm{eff}(z)$ in Fig.~\ref{Fig:FigSup7}(a) with $\nu=0.001$. The propagation dynamics obtained from Eq.~(\ref{Eq:NSE}) is shown in Fig.~\ref{Fig:FigSup13}(a1). Not surprisingly, the corresponding center-of-mass trajectory exhibits quantized motion [Fig.~\ref{Fig:FigSup13}(a2)]. 

Then, we proceed to consider a case where we slowly ramp $g(z)$ from an initial value $g_0$ to the final value $g_f$ in a passage according to 
\begin{equation}\label{gz}
	g(z) = g_0 + (g_f - g_0)\left(1 - e^{-z/\sigma}\right),
\end{equation}
where $\sigma$ controls the ramp velocity. This corresponds to a slow quench of the nonlinear Hermitian Hamiltonian. Choosing $g_0 = 1$, $g_f = 0.254$, and $\sigma = 2\pi$, we find quantized displacement despite this aperiodic time variation of nonlinearity, as shown in Figs.~\ref{Fig:FigSup13}(b1) and (b2). Note, however, our numerical results suggest that the trapped or pumped soliton is not, in general, tied to the instantaneous nonlinear Hamiltonian at the end of a pump cycle, and quantized displacement may not be guaranteed for an arbitrary time-dependent nonlinearity. A more systematic exploration as to what kind of aperiodically varying nonlinearity leads to quantized displacement is beyond the scope of this paper and will be presented in a future work.